\documentclass[12pt,preprint]{aastex}
\usepackage[onecolumn]{emulateapj5}

\newcommand{\gtsima}{$\; \buildrel > \over \sim \;$}
\newcommand{\ltsima}{$\; \buildrel < \over \sim \;$}
\newcommand{\simgt}{\lower.5ex\hbox{\gtsima}}
\newcommand{\simlt}{\lower.5ex\hbox{\ltsima}}
\newcommand{\himpc}{{\hbox {$\,h^{-1}$}{\rm Mpc}} }

\newcommand{\bfm}[1]{{\mbox{\boldmath $#1$}}}
\newcommand{\sbfm}[1]{{\mbox{\scriptsize\boldmath $#1$}}}
\newcommand{\mTheta}{{\mit \Theta}}

\newcommand{\mPsi}{{\mit \Psi}}
\newcommand{\mPhi}{{\mit \Phi}}
\newcommand{\mOmega}{{\mit \Omega}}
\newcommand{\mGamma}{{\mit \Gamma}}

\begin{document}

\renewcommand{\theequation}{\mbox{\rm
{\arabic{section}.\arabic{equation}}}} 


\title{Eigenmode Analysis of Galaxy Distributions in Redshift Space}

\author{Takahiko Matsubara}
\affil{Department of Physics and Astrophysics, 
	Nagoya University,
	Chikusa, Nagoya 464-8602, Japan}

\author{Alexander S. Szalay, Adrian C. Pope}
\affil{Department of Physics and Astronomy, 
        The Johns Hopkins University,
        Baltimore, MD 21218}

\begin{abstract}

Eigenmode analysis is one of the most promising methods of analyzing
large data sets in ongoing and near-future galaxy surveys. In such
analyses, a fast evaluation of the correlation matrix in arbitrary
cosmological models is crucial. The observational effects, including
peculiar velocity distortions in redshift space, light-cone effects,
selection effects, and effects of the complex shape of the survey
geometry, should be taken into account in the analysis. In the
framework of the linear theory of gravitational instability, we
provide the methodology to quickly compute the correlation matrix. Our
methods are not restricted to shallow redshift surveys, arbitrarily
deep samples can be dealt with as well. Therefore, our methods are
useful in constraining the geometry of the universe and the dark
energy component, as well as the power spectrum of galaxies, since
ongoing and near-future galaxy surveys probe the universe at
intermediate to deep redshifts, $z \sim$ 0.2--5. In addition to the
detailed methods to compute the correlation matrix in 3-dimensional
redshift surveys, methods to calculate the matrix in 2-dimensional
projected samples are also provided. Prospects of applying our methods
to likelihood estimation of the cosmological parameters are discussed.

\end{abstract}



\setcounter{equation}{0}
\section{
Introduction
\label{sec1}
}

The large-scale structure (LSS) of the universe offers invaluable
information on cosmology. The clustering pattern of galaxies reflects
what the primordial universe looks like and how the universe evolves.
In contrast to the observations of the cosmic microwave background
(CMB), in which only information from photons emitted at a fixed
redshift $z_{\rm dec} \simeq 1090$ are observed, the observations of
the LSS provide information from photons emitted at various redshifts
$z \simlt 5$. Therefore, while the CMB primarily tells us the state of
the early universe, the LSS shows us that of the present universe and
its recent evolution. Detailed analysis of the LSS is extremely
important in order to obtain a consistent picture of the evolving
universe.

The spatial distribution of galaxies has particular importance in
cosmology. Traditionally the power spectrum, or the correlation
function is computed from galaxy distributions and is compared with
theoretical predictions. Strictly speaking, the power spectrum and the
correlation function are properly evaluated only when the sample is
homogeneous. In a real-world observation, homogeneous sampling of
galaxies is not possible. The sampling density varies with radial
distance, and possibly with survey directions. The survey geometry may
have a complex shape. The observed galaxies are inevitably on a
light-cone, therefore evolutionary effect comes in when a survey has a
certain depth.

Therefore, estimating the power spectrum from a real data needs a
method to approximately erase those sources of inhomogeneity. Among
others, \citet{fel94} offers a standard method to estimate the power
spectrum. The estimated power spectrum by such standard method is a
convolution of the true power with the window function of the survey.
This becomes a problem when the survey has a complex shape.

In a seminal paper, \citet{vog96} introduced a novel method, i.e., the
method of Karhunen-Lo\`eve (KL) eigenmodes in the context of galaxy
redshift surveys. This method can handle complex shape of the survey
geometry as well as complex selection functions. In the context of CMB
analysis, \citet{bon95} introduced an equivalent method which is
called signal-to-noise eigenmodes method. The KL method is now
recongnized as one of the most promising methods to tackle large data
sets of the cosmological surveys \citep{teg97,teg98,tay01}, and has
been successfully applied to the Las Campanas Redshift Survey (LCRS)
\citep{mat00b}, the Sloan Digital Sky Survey (SDSS)
\citep{sza03,pop04}, the ROSAT-ESO Flux-Limited X-Ray (REFLEX) Galaxy
Cluster Survey \citep{sch02}. A variant, simplified method, which is
called pseudo-KL eigenmodes method is applied to the IRAS Point Source
Catalog Redshift Survey (PSC$z$) \citep{teg00}, the 2dF Galaxy
Redshift Survey \citep{teg02}, and SDSS \citep{teg03}.

In applying the KL technique to cosmology, it is important to have a
good method to theoretically compute the power spectrum of the KL
modes. This is not a trivial task, since the observed distribution of
the galaxies are distorted by observational effects including peculiar
velocity distortions, light-cone effects, varying selection functions,
and complex shape of the survey geometry. Among these, the light-cone
effect has never been taken into account in previous applications of
the KL method to redshift surveys. The light-cone effects consist of
evolutionary effects and geometric effects. The clustering of galaxies
evolves with redshift through evolution in mass clustering, and
evolution in the biasing of galaxies. The comoving distance is not
proportional to the redshift and the apparent clustering pattern in
redshift space is not the same as in comoving space. For shallow
redshift surveys in which the redshift range is, e.g., $0 \simlt z
\simlt 0.2$, the light-cone effect has small contributions. However,
QSO surveys probe much deeper regions, and ongoing and near-future
redshift surveys, like the Deep Extragalactic Evolutionary Probe
(DEEP) survey \citep[e.g.,][]{dav03}, the SDSS Luminous Red Galaxy
(LRG) survey \citep{eis01}, and the Kilo-Aperture Optical Spectrograph
(KAOS) project\footnote{http://www.noao.edu/kaos/} probe the universe
of redshifts $0.2 \simlt z \simlt 5$.

Therefore, it is crucial in cosmological analyses of large surveys to
have a good methodology in applying the KL eigenmodes method to
ongoing and near-future redshift surveys which are not necessarily
shallow. In this paper, methods to efficiently evaluate the
theoretical values of the correlation matrix, playing a central role
in the KL analysis, are developed and described in detail. This paper
is organized as follows. In \S 2 the Bayesian analysis of the large
data set in the context of galaxy surveys is reviewed, and necessity
of a fast method to compute the correlation matrix is explained. In \S
3 the central method of this paper is presented. Necessary quantities
in describing the method are introduced in \S 3.1. A fast method to
compute the linear correlation matrix in redshift space is given in \S
3.2 and \S 3.3. The nonlinear Finger-of-God effect is discussed in \S
3.4. Fast methods to compute the linear correlation matrix in
projected 2-dimensional samples are provided in \S 3.5. In \S 4
additional techniques in applying the present methods to the KL
eigenmodes are explained. Finally, the presented methods are
summarized and discussed in \S 5.

\setcounter{equation}{0}
\section{
The Eigenmode Analysis and Importance of the Correlation Matrix
\label{sec2}
}

In this section we briefly review the eigenmode analysis originally
introduced by \citet{vog96} in the context of the redshift surveys.
The importance of accurately modeling the correlation matrix in
redshift space and of developing a numerically fast algorithm to
calculate the model matrices, which are the main subject of this work,
will be clarified in this section.

\subsection{Bayesian Analyses of the Huge Data Sets
\label{sec2-1}}

One of the central motivation of the eigenmode analysis comes from the
difficulty in direct Bayesian analysis of huge data sets of observed
galaxy distributions. It is desirable that the cosmological models are
discriminated by the galaxy distribution itself, not relying on the
inverse problem of determining the underlying density power spectrum
from observed galaxy distributions. While a cosmological model
predicts the linear power spectrum, the dependence of the galaxy
clustering in redshift surveys on the theoretical power spectrum is
not simple.

Nonlinearity in the gravitational evolution distorts the power
spectrum on small scales $\simlt 10\himpc$. Peculiar velocities
anisotropically distort the power spectrum on all scales. Spatial
variation of the mass-to-light ratio or the galaxy biasing distorts
the power spectrum. In deep redshift surveys, the spatial curvature
and the evolution of the Hubble parameter distort the apparent pattern
of the galaxy clustering. These various distortions in the observed
power spectrum also depend on the cosmological model. Even if one can
invent and adopt some ansatz on such a complex inversion problem, fair
amount of information will be lost in this inversion process.

However, the reconstruction of the underlying power spectrum is not
necessary to discriminate the cosmological models. In the Bayesian
approach, the posterior probability of the set of model parameters
$\mTheta = \{\theta_\alpha\}$, given a set of observational data $D =
\{d_i\}$ and additional prior information $I$, is estimated by Bayes's
theorem:
\begin{equation}
   P\left(\mTheta|D,I\right) =
   P\left(\mTheta|I\right) 
   \frac{P\left(D|\mTheta,I\right)}
      {P\left(D|I\right)}.
\label{eq2-1}
\end{equation}
The first factor is the Bayesian prior, i.e., the prior knowledge on
the model parameters, and the denominator is the normalization
constant. Evaluating the numerator is the crucial part of the Bayesian
approach. One needs to repeatedly evaluate this factor to maximize the
posterior probability in multi-dimensional parameter space. Therefore,
the numerator should be numerically calculated as fast as possible.
This part of the calculation is usually not easy, especially when the
the data $D$ consists of a large set of observational figures. In
galaxy redshift surveys, the data $D$ consists of the distribution of
galaxies in redshift space. These days the scales of redshift surveys
become more and more immense. In the 2dF Galaxy Redshift Survey
\citep{col01}, redshifts of about 220,000 galaxies are observed in the
sky coverage of 2,000 square degrees. In the Sloan Digital Sky Survey
(SDSS) \citep{yor00}, redshifts of about 1,000,000 galaxies are planned
to be observed over 10,000 square degrees of the sky.

When the prior information on the model, $I$, has nothing to do with
obtaining the data themselves, the numerator of equation (\ref{eq2-1})
is equal to $P(D|\mTheta)$. In addition, when the statistical
distribution of the data $D$ is given by the multivariate Gaussian
distribution, this factor has the form,
\begin{equation}
   P(D|\mTheta) =
   \frac{1}{\sqrt{(2\pi)^N \det R(\mTheta)}}
   \exp\left[ -\frac12
     \left(\bfm{d}^{\rm T} - \langle\bfm{d}^{\rm T}\rangle_{\mTheta}\right)
       R^{-1}(\mTheta) \left(\bfm{d} -
   \langle\bfm{d}\rangle_{\mTheta}\right)
   \right],
\label{eq2-2}
\end{equation}
where $\bfm{d}$ is the data vector $\bfm{d} = (d_1, d_2, \cdots,
d_N)^{\rm T}$ and $R$ is the $N\times N$ correlation matrix given by
\begin{equation}
  R(\mTheta) =
    \left\langle
      \left(\bfm{d} - \left\langle \bfm{d}
        \right\rangle_\mTheta\right) 
      \left(\bfm{d} - \left\langle \bfm{d}
        \right\rangle_\mTheta\right)^{\rm T}
    \right\rangle_{\mTheta} =
    \left\langle \bfm{d} \bfm{d}^{\rm T} \right\rangle_{\mTheta} -
    \left\langle \bfm{d} \right\rangle_{\mTheta}
    \left\langle \bfm{d}^{\rm T} \right\rangle_{\mTheta}.
\label{eq2-3}
\end{equation}
In the above notations, $\langle\cdots\rangle_\mTheta$ indicates the
ensemble average over the statistical realizations of the universe
\citep[see, e.g., ][]{pee80} for an assumed set of model parameters
$\mTheta$. The correlation matrix therefore explicitly depends on the
model parameters.

In the analysis of redshift surveys, the observed set of data is
primarily a set of the positions of galaxies in redshift space. The
density field is derived from this primary data, which is one of the
most fundamental information of our universe in redshift surveys.
Therefore, it is natural that the data $D$ is identified as the number
density field of galaxies in redshift space. For this purpose, the
redshift space is pixelized to obtain a discretized set of galaxy
counts. Thus, the set of the number counts $d_i$ in a cell $i$ is the
data $D$ in redshift surveys. The pixel size determines the resolution
in the analysis, so it is desirable to adopt as many cells in redshift
space as possible. However, too many cells cause trouble in
calculating the inverse of the correlation matrix in equation
(\ref{eq2-2}). The caluculation of the inverse of the huge matrix
requires a lot of computation time, which is primarily an $N^3$
process and can be an $N^{2.8}$ process at most \citep{pre92}. This is
the bottle neck in the Bayesian analysis since the inversion should be
repeatedly performed for every set of parameters.

The pixelized raw data vector $\bfm{d} = (d_1, d_2, \ldots, d_N)^{\rm
T}$ consists of clustering signals and noises. Most of the process in
inverting the correlation matrix in the above procedure might be
devoted to dealing with noises. Since the information on the
cosmological model is only contained in the signals, it is beneficial
to reduce the dimension of the correlation matrix, keeping the maximal
information on the clustering signals and minimizing contributions by
the noise. One simple way to achieve this aim is to consider a linear
projection of the raw data vector $\bfm{d}$ into vector $\bfm{B}$ of a
smaller dimension:
\begin{equation}
   \bfm{B} = P \bfm{d},
\label{eq2-4}
\end{equation}
where the projection matrix $P$ has a dimension $M\times N$, and
the dimension of the vector $\bfm{B}$ is $M$, where $M < N$. The
correlation matrix of the reduced data vector is given by
\begin{equation}
  C(\mTheta) = 
  \left\langle
    \left(\bfm{B} - \left\langle \bfm{B}
      \right\rangle_\mTheta\right) 
    \left(\bfm{B} - \left\langle \bfm{B}
      \right\rangle_\mTheta\right)^{\rm T}
  \right\rangle_{\mTheta} =
  P R(\mTheta) P^{\rm T}.
\label{eq2-5}
\end{equation}
The linear transformation from $\bfm{d}$ to $\bfm{B}$ ensures the
preservation of the Gaussianity of the distribution.

A clever choice of the projection matrix $P$ enables us to avoid the
inversion of the originally huge correlation matrices without losing
most of the information contained in signals. In such a case, the
inversion of smaller correlation matrices for the reduced data
suffices for the Bayesian analysis, and the computation time is
reduced by a factor of about $(M/N)^{2.8}$. In Appendix~\ref{app1},
the construction of the projection matrix by the KL eigenmodes with
maximal signal-to-noise ratio is comprehensively reviewed. In short,
the projection matrix by KL eigenmodes picks the first $M$ modes that
have largest S/N ratio.

The probability of having a reduced data set, $D_{\rm reduced} =
\{B_n\}$, given the set of model parameters, $\mTheta$, is therefore
given by
\begin{equation}
   P(D_{\rm reduced}|\mTheta) =
   \frac{1}{\sqrt{(2\pi)^M \det C(\mTheta)}}
   \exp\left[ -\frac12
     \left(\bfm{B}^{\rm T} - \langle\bfm{B}^{\rm T}\rangle_{\mTheta}\right)
       C^{-1}(\mTheta) \left(\bfm{B} -
   \langle\bfm{B}\rangle_{\mTheta}\right)
   \right].
\label{eq2-0}
\end{equation}
The above equation assumes that the procedure of obtaining data is
independent of the assumed model parameters, so that the projection
matrix should be fixed throughout analyzing and maximizing the
likelihood function. However, the optimal choice of the projection
matrix usually depends on the underlying correlation function, and
thus on model parameters, as explicitly presented in
Appendix~\ref{app1} in the case of the KL-mode projection. Therefore,
the projection matrix is initially constructed by a fiducial set of
model parameters, which is fixed throughout the maximization of the
likelihood function.

While the reduced correlation matrix $C$ constructed by KL eigenmodes
is diagonalized for a fiducial set of model parameters, it should be
noted that the correlation matrix $C$ is {\em not} diagonal for other
set of model parameters. Therefore, the inversion of the correlation
matrix $C$ still needs to be performed. The likelihood function is not
biased by the initial choice of the fiducial model, because the
fiducial parameters only determine the projection matrix $P$ which is
fixed throughout. If the initial choice of the fiducial model is far
from the correct model, the projection matrix is not the optimal one,
and therefore we obtain rather broader profile of the likelihood
function around the maximum, because of picking low-S/N modes. One can
iterate the likelihood maximization by choosing a fiducial model from
the preceding estimation to acquire the narrower profile of the
likelihood function. This is the procedure we apply in practical
situations.

The model correlation matrix $C(\mTheta)$ of a reduced data set should
be repeatedly calculated for various models, fixing the projection
matrix $P$. One still needs to construct the $N\times N$ matrix
$R(\mTheta)$ as many times as $C$, because $C(\mTheta) = P R(\mTheta)
P^{\rm T}$. It is only when this process is omitted that the parameter
dependence on $R$ is linear or polynomial. However, most of the
important parameters are not linearly or polynomially dependent on the
correlation matrix. On that ground, a good modeling and numerically
fast construction of the correlation matrix $R$ is essential.

\subsection{
Bayesian Analysis of the Galaxy Distribution in Redshift Surveys
\label{sec2-2}}

Now we turn our attention to the application of the general method
explained above to the analysis of the data in galaxy redshift
surveys. The primary data set in redshift surveys is the galaxy
distribution in redshift space. In order to directly deal with the
galaxy distribution, first we pixelize the redshift space, and then
build the data vector $\bfm{d}$ from galaxy counts $d_i$ in cells of
the volume $V_i$. The expected counts $N_i = \langle d_i \rangle$ are
given by an integral of the selection function $n(\bfm{s})$, i.e.,
expected number density field in a survey:
\begin{equation}
   N_i =
   \int d^3s K_i(\bfm{s}) n(\bfm{s}),
\label{eq2-7}
\end{equation}
where $K_i(\bfm{s})$ is the kernel function of a cell $i$, which is
centered at $\bfm{s}_i$, and normalized by $\int d^3s K_i(\bfm{s}) =
V_i$. For example, the top-hat kernel of smoothing radius $R$ is given
by $K_i(\bfm{s}) = \mTheta(R-|\bfm{s}-\bfm{s}_i|)$ and $V_i = 4\pi
R^3/3 $, where $\mTheta(x)$ is the Heaviside step function and
$\bfm{s}_i$ is the center of the cell $i$. We require the knowledge of
the selection function in a survey. The correlation matrix is given by
the moments of the count-in-cells \citep{pee80}, and is related to the
convolution of the two-point correlation function in redshift space,
$\xi^{\rm (s)}(\bfm{s},\bfm{s}')$. Including the shot noise term, the
relation is
\begin{equation}
   R_{ij} = \int d^3s d^3s' K_i(\bfm{s}) K_j(\bfm{s}')
   n(\bfm{s}) n(\bfm{s}') \xi^{\rm (s)}(\bfm{s},\bfm{s}') +
   \int d^3s K_i(\bfm{s}) K_j(\bfm{s}) n(\bfm{s}) + E_{ij},
\label{eq2-8}
\end{equation}
where the first term corresponds to the signal, second term
corresponds to the shot noise, and the last term, $E_{ij}$, is the
correlation matrix for other sources of noise such as magnitude errors
or uncertainty in the luminosity function, which is assumed to be
independent from the clustering signal for simplicity. One can
reasonably assume that the selection function $n(\bfm{s})$ within each
cell is approximately constant and can be replaced by $n_i = N_i/V_i$,
in which case the equation (\ref{eq2-8}) reduces to
\begin{equation}
   R_{ij} = N_i N_j \xi^{\rm (s)}_{ij} +
   \sqrt{N_i N_j} K_{ij} + E_{ij},
\label{eq2-9}
\end{equation}
(no sum over $i$, $j$) where
\begin{equation}
   \xi^{\rm (s)}_{ij} =
   \frac{1}{V_i V_j} 
   \int d^3s d^3s'
   K_i(\bfm{s}) K_j(\bfm{s}')
   \xi^{\rm (s)}\left(\bfm{s}, \bfm{s}'\right),
\label{eq2-9-1}
\end{equation}
and 
\begin{equation}
   K_{ij} =
   \frac{1}{\sqrt{V_i V_j}} 
   \int d^3s
   K_i(\bfm{s}) K_j(\bfm{s}).
\label{eq2-9-2}
\end{equation}
The first term in equation (\ref{eq2-9}) contains the model
parameters, and corresponds to the signal part of the correlation
matrix, $S_{ij}$. The second term corresponds to the shot noise. When
the cells do not overlap, the matrix $K_{ij}$ is diagonal:
\begin{equation}
   K_{ij} =
   \frac{\delta_{ij}}{V_i} 
   \int d^3s
   {K_i}^2(\bfm{s}), \qquad \mbox{(non-overlapping)},
\label{eq2-9-3}
\end{equation}
(no sum over $i$, $j$), where $\delta_{ij}$ is the Kronecker's delta.
For the non-overlapping top-hat kernel, this equation is simplified as
$K_{ij} = \delta_{ij}$, and the correlation matrix is simply given by
\begin{equation}
   R_{ij} = N_i N_j \xi^{\rm (s)}_{ij} +
   N_i \delta_{ij} + E_{ij}, \qquad (\mbox{non-overlapping,\ top-hat})
\label{eq2-9-4}
\end{equation}
(no sum over $i$, $j$).

Unlike the correlation function in real space, the correlation
function in redshift space is neither isotropic nor homogeneous. The
correlation function in real space is a function of separations
between two-points, $\xi(|\bfm{x}_1 - \bfm{x}_2|)$. However, the
redshift space is distorted from the real space, and consequently the
correlation function is no longer the function of the separation only.

The comoving distance to a galaxy is not a primary observable quantity
and the redshift is the only estimate of the real distance from the
Earth to the cosmologically distant galaxies. Since the redshift is
not identical to the comoving distance, the redshift space is
anisotropically distorted from the real space. There are two
predominant sources for the redshift-space distortion.

The first source of the distortion is the peculiar velocities of
galaxies. The peculiar velocities along the line of sight increase or
decrease the observed redshift by the Doppler effect. On nonlinear
scales, the peculiar velocities are dominated by the random motion in
a cluster potential. On linear scales, the coherent motion along the
density gradient is predominant. The linear velocity distortions are
relatively simple when the distant-observer (or plane-parallel)
approximation is applied. \citet{kai87} found the linear power
spectrum in redshift space, $P^{\rm (s)}(\bfm{k})$ is anisotropically
distorted along the line of sight, and is given by the formula,
\begin{equation}
   P^{\rm (s)}(\bfm{k}) = (1 + \beta \mu_\sbfm{k}^2)^2 P(k),
\label{eq2-10}
\end{equation}
where $P(k)$ is the power spectrum in real space, $\mu_\sbfm{k}$ is
the cosine of the wavevector $\bfm{k}$ to the line of sight. The
counterpart of this formula for the linear two-point correlation
function, $\xi^{\rm (s)}(\bfm{r})$, which is given by the Fourier
transform of equation (\ref{eq2-10}), was derived by \citet{ham92}.

The second source of the distortion is what is called
Alcock-Paczy\'nski (AP, hereafter) effect. \citet{alc79} pointed out
that the spherical objects in deep redshift region appears to be
distorted in redshift space because of the nonlinear relation between
the redshift and the comoving distance, and of the nonlinear relation
between the apparent angular size and the angular diameter distance.
Suppose that there is a spherical object of diameter $l$ in units of
comoving length. The redshift difference of this object is $\Delta z =
H(z) l/c$, and the apparent angular size is $\Delta\theta = l/D_{\rm
A}(z)$, where $H(z)$ is the redshift-dependent Hubble parameter and
$D_{\rm A}(z)$ is the comoving angular diameter distance [proper
angular diameter distance is given by $d_{\rm A}(z) = D_{\rm
A}(z)/(1+z)$]. The ratio
\begin{equation}
   \frac{\Delta z}{z\Delta\theta} = \frac{D_{\rm A}(z)}{cz/H(z)}
\label{eq2-11}
\end{equation}
is unity for sufficiently small $z$. In deep redshift regions,
however, the ratio significantly deviates from unity and the deviation
pattern strongly depends on the density parameter $\mOmega_{\rm M0}$
and on the cosmological constant $\mOmega_{\Lambda 0}$, or parameters
of the dark energy.

The redshift-space distortions of the two kinds, i.e., the velocity
distortions and the AP distortions, are interwoven in the real
universe except for the shallow redshift surveys without the AP
effect. \citet{bal96} derived the AP effect on the linear power
spectrum, and \citet{mat96} derived the same effect on the linear
two-point correlation function, both using the distant observer
approximation. Modern redshift surveys havee survey geometries with
wide opening angles. In these cases, the distant-observer
approximation is inappropriate and the wide-angle effect is not
negligible in order to handle the information contained in the survey.
\citet{sza98} derived a formula of velocity distortions in the linear
two-point correlation function with wide-angle effect. \citet{mat00}
derived the most general formula of the linear two-point correlation
function in which the velocity distortions, the AP effect, and the
wide-angle effect are all included in a unified formula. All the
previous formulas for the linear two-point correlation function are
limiting cases of the last formula. Our construction of the
correlation matrix $R$, which is explained in the following section,
takes full advantage of the last expression.


\setcounter{equation}{0}
\section{
Constructing Correlation Matrices in Redshift Space
\label{sec3}
}

\subsection{
The General Formula of the Correlation Function in Redshift Space
}

To obtain the correlation matrix of equation (\ref{eq2-9}), it is
necessary to have the analytic expression of the correlation function
$\xi^{\rm (s)}(\bfm{s},\bfm{s}')$, which is given by \citet{mat00}.
For completeness, we reproduce the formula in this section. Before
that, we need to introduce some basic notations.

The unperturbed metric is given by the Robertson-Walker (RW) metric,
\begin{equation}
  ds^2 = - dt^2 +
  a^2(t) \left[
    dx^2
    + {S_K}^2(x) \left(d\theta^2 + \sin^2\theta \phi^2\right)
  \right],
\label{eq3-1}
\end{equation}
where we employ the unit system with $c=1$, and adopt a notation,
\begin{equation}
  S_K(x) =
  \left\{
  \begin{array}{ll}
    \displaystyle
    \frac{{\rm sinh}\left(\sqrt{-K} x\right)}{\sqrt{-K}}, &
    (K < 0), \\
    x, & (K = 0), \\
    \displaystyle
    \frac{\sin\left(\sqrt{K} x\right)}{\sqrt{K}}, &
    (K > 0),
  \end{array}
  \right.
\label{eq3-2}
\end{equation}
The comoving angular diameter distance is related to the last function
by
\begin{equation}
   D_{\rm A}(z) = S_K\left(x(z)\right).
\label{eq3-3}
\end{equation}
where $z$ is the ``RW redshift'' in which the peculiar velocity effect
is not included, i.e., the redshift of the unperturbed universe. The
radial coordinate $x$ is the comoving distance from the observer, and
the relation to the RW redshift is given by
\begin{equation}
   x(z) = \int_0^z  \frac{dz'}{H(z')},
\label{eq3-4}
\end{equation}
where $H(z) = \dot{a}/a$ is the time-dependent Hubble parameter. We
consider the general case that the dark energy has an equation of
state $p(z) = w(z) \rho(z)$, in which case,
\begin{equation}
   H(z) = H_0
   \sqrt{
      (1+z)^3 \mOmega_{\rm M0} - 
      (1+z)^2 \mOmega_{\rm K0} +
      (1+z)^3 \exp\left(3\int_0^z \frac{w dz}{1+z}\right) \mOmega_{\rm Q0}
   },
\label{eq3-5}
\end{equation}
where $\mOmega_{\rm M0}$ is the density parameter of matter(s),
$\mOmega_{\rm Q0}$ is the density parameter of the dark energy, and
$\mOmega_{\rm K0} = \mOmega_{\rm M0} + \mOmega_{\rm Q0} - 1$ is the
curvature parameter.

The linear growth factor $D(z)$ is a growing solution of the ordinary
differential equation,
\begin{equation}
   \ddot{D} + 2 H(z) \dot{D} - \frac{3}{2} {H_0}^2 \mOmega_{\rm M0} (1 +
   z)^3 D = 0,
\label{eq3-6}
\end{equation}
where the dot represents the differentiation with respect to the
proper time $t$, and the proper time $t$ is related to the RW
redshift $z$ by
\begin{equation}
   t = \int_z^\infty \frac{dz'}{(1 + z')H(z')}.
\label{eq3-7}
\end{equation}
It is useful to transform the equation (\ref{eq3-6}) to the following
set of equations \citep{mat03},
\begin{eqnarray}
&&
  \frac{d\ln D}{d\ln a} = f,
\label{eq3-8a}\\
&&
  \frac{df}{d\ln a} = -f^2 - \left(1 - \frac{\mOmega_{\rm M}}{2} -
  \frac{1 + 3 w}{2} \mOmega_{\rm Q} \right) f +
  \frac{3}{2} \mOmega_{\rm M}, 
\label{eq3-8b}
\end{eqnarray}
where $a = (1 + z)^{-1}$ is the scale factor of the universe, and
\begin{eqnarray}
&&
  \mOmega_{\rm M}(z) = \frac{{H_0}^2}{H^2(z)} (1+z)^3 \mOmega_{\rm M0}
\label{eq3-9a}\\
&&
  \mOmega_{\rm Q}(z) =
  \frac{{H_0}^2}{H^2(z)}
  \exp\left(3\int_0^z \frac{1+w}{1+z} dz\right)
  \mOmega_{\rm Q0}
\label{eq3-9b}
\end{eqnarray}
are the time-dependent density parameters of matter and dark energy,
respectively. Determining both the growth factor, $D(z)$, and the
logarithmic derivative of the growth factor, $f(z)$, is neccesary in
the formula of the redshift-space distortions. Accordingly, the
Runge-Kutta integrations of the set of equations (\ref{eq3-8a}) and
(\ref{eq3-8b}) meet the requirement.

The linear formula for the two-point correlation function in redshift
space is presented in Appendix~\ref{app2}. It has the form,
\begin{equation}
  \xi^{\rm (s)}(z_i,z_j,\theta_{ij}) = 
  b_i b_j D_i D_j
  \sum_{n=0}^2 \sum_{l=0}^{2n}
  c_l^{(n)}(x_i,x_j,\theta_{ij}) \xi_l^{(n)}(x_{ij}),
\label{eq3-10}
\end{equation}
where $z_i$ and $z_j$ are the redshifts of the two-points and
$\theta_{ij}$ is the angle between them. The linear bias parameter
$b(z)$ as a function of the redshift is introduced. The distances $x_i
= x(z_i)$, $x_j = x(z_j)$ are comoving distances to the two points
according to equation (\ref{eq3-4}). The comoving separation $x_{ij}$
between the two points is calculated by equation (\ref{eqa3-19}). The
functions $c_l^{(n)}$ are given by geometrical quantities and defined
by equations (\ref{eqa3-20a})--(\ref{eqa3-20i}). The functions
$\xi_l^{(n)}$ are given by integrations of the power spectrum, and
defined by equation (\ref{eqa3-17}). Although the expression
(\ref{eq3-10}) is somewhat long, it is straightforwardly calculated
once the single-variable functions $f(z)$, $D(z)$, $x(z)$, $b(z)$,
$\xi_l^{(n)}(x)$ are calculated and tabulated beforehand for a fixed
cosmological model.

While the correlation function linearly depends on the power spectrum
through the function $\xi_l^{(n)}(x)$, the dependences on other
cosmological parameters like $\mOmega_{\rm M0}$, $\mOmega_{\rm Q0}$,
$w(z)$ are nonlinear. The bias parameter $b(z)$ only appears as
2nd order polynomials


\subsection{
Smoothing Integrations of the Correlation Function
}

Evaluating the correlation matrix of the equation (\ref{eq2-9})
requires the six-dimensional integration of the correlation function
in general. Direct integration is not a good idea. It is extremely
important to obtain the elements of the correlation matrix without
involving time-consuming steps such as multi-dimensional numerical
integrations, because we need to construct many huge correlation
matrices, with various cosmological models. We need to develop a
method to calculate them with algebraic and interpolating methods
without numerical integration for each element. There is actually such
a method with special choices of the shape of the smoothing kernel as
shown below.

The signal part of the correlation matrix of equation (\ref{eq2-9-1})
is essentially the correlation function of the smoothed density
contrast,
\begin{equation}
  \xi^{\rm (s)}_{ij} =
  \left\langle
    \delta^{(s)}_{R}(x_i,\theta_i,\phi_i)
    \delta^{(s)}_{R}(x_j,\theta_j,\phi_j)
  \right\rangle
\label{eq3-20}
\end{equation}
where 
\begin{equation}
  \delta^{\rm (s)}_{R}(x,\theta,\phi) =
  \int {S_K}^2(x') dx' \sin\theta' d\theta' d\phi'
  W_{R}(x,\theta,\phi;x',\theta',\phi')
  \delta^{\rm (s)}(x',\theta',\phi')
\label{eq3-21}
\end{equation}
is the smoothed density contrast in redshift space, which is convolved
by an normalized kernel function $W_R$. The normalization of the
kernel is given by
\begin{equation}
  \int {S_K}^2(x') dx' \sin\theta' d\theta' d\phi'
  W_{R}(x,\theta,\phi;x',\theta',\phi') = 1.
\label{eq3-21-1}
\end{equation}
The kernel function $W_R(x,\theta,\phi;x',\theta',\phi')$ is non-zero
only when the comoving separation between $(x,\theta,\phi)$ and
$(x',\theta',\phi')$ is comparable to, or less than $R$. Practically,
$R$ is much less than the Hubble scale, and the linear operator in the
equation (\ref{eqa3-16-1}) is commutable with the smoothing
integration of equation (\ref{eq3-21}), since the functions $D(z)$,
$b(z)$, $f(z)$, $\alpha(z)$, and ${S_K}(x)$ are practically constants
over individual cells, and the Laplacian is an Hermitian operator.
Therefore, the equation (\ref{eq3-21}) reduces to
\begin{equation}
  \delta^{\rm (s)}_{R}(x(z),\theta,\phi) =
  D(z)
  \left\{
    b(z) + f(z)
      \left[\frac{\partial}{\partial x} + \alpha(z)\right]
        \frac{\partial}{\partial x} (\triangle + 3K)^{-1}
  \right\}
  \delta_R(x(z),\theta,\phi),
\label{eq3-22}
\end{equation}
where 
\begin{equation}
  \delta_R(x,\theta,\phi) =
  \int {S_K}^2(x') dx' \sin\theta' d\theta' d\phi'
  W_{R}(x,\theta,\phi;x',\theta',\phi')
  \delta(x',\theta',\phi'),
\label{eq3-23}
\end{equation}
is the smoothed density contrast in real space. A complex shape of the
smoothing kernel complicates further analytic treatment, and therefore
slows down the numerical evaluation of the correlation matrix based on
the analytic formula. It is desirable to choose the smoothing kernel
as simple as possible. In this repect, the simplest choice of the
smoothing function $W_R$ is the spherically symmetric function in
comoving space, in which case, the kernel is expanded by complete set
of orthonormal modes $X_0$ (c.f., eq.[\ref{eqa2-6b}]),
\begin{equation}
  W_R(x_{\rm rel}) =
  \int \frac{k^2dk}{2\pi^2} W(kR) X_0(k,x_{\rm rel}),
\label{eq3-24}
\end{equation}
where $x_{\rm rel}$ is the relative separation in comoving units
between $(x,\theta,\phi)$ and $(x',\theta',\phi')$. The
equation (\ref{eq3-24}) is inverted as
\begin{equation}
  W(kR) = 4\pi \int {S_K}^2(x) dx W_R(x) X_0(k,x).
\label{eq3-25}
\end{equation}
Since $R$ is much less than the curvature scale, $R \ll |K|^{-1/2}$,
the above equation accurately approximates to 
\begin{equation}
  W(kR) = 4\pi \int x^2 dx W_R(x) j_0(kx) =
  \int d^3x W_R(x) e^{-i\sbfm{k}\cdot\sbfm{x}}
\label{eq3-26}
\end{equation}
Thus, the function $W(kR)$ is the Fourier transform of the kernel, or
the spherical window function of the kernel. The popular choices are
the top-hat window function $W(kR) = 3 j_1(kR)/(kR) = 3(\sin kR - kR
\cos kR)/(kR)^3$, and the Gaussian window function $W(kR) = \exp(-k^2
R^2/2)$. Useful window functions are given in Appendix~\ref{app3},
including the $m$-weight Epanechnikov kernel, which might also be
useful.

There is a useful addition theorem for $X_l$ \citep[e.g.,][]{rat95,mat00},
\begin{equation}
  X_0(k,x_{\rm rel}) =
  \sum_{l=0}^\infty (2l+1) X_l(k,x) X_l(k,x') P_l(\cos\theta_{\rm rel}),
\label{eq3-27}
\end{equation}
where $X_l$ is given by equation (\ref{eqa2-4}), $P_l(x)$ are the
Legendre polynomials, $\theta_{\rm rel}$ is the relative angle between
the directions $(\theta,\phi)$ and $(\theta',\phi')$. The sum rule
(\ref{eq3-27}) holds even for a non-flat case $K \neq 0$ as well as a
flat case $K=0$. The Legendre polynomials are also expanded by the
spherical harmonics $Y_l^m$,
\begin{equation}
  P_l(\cos\theta_{\rm rel}) =
  \frac{4\pi}{2l+1} \sum_{m=-l}^l
   Y_l^m(\theta,\phi) {Y_l^m}^*(\theta',\phi').
\label{eq3-28}
\end{equation}
Combining equations (\ref{eq3-24}), (\ref{eq3-27}), and
(\ref{eq3-28}), the spherical kernel function $W_R$ is explicitly
represented by the arguments $(x,\theta,\phi;x',\theta',\phi')$ to be
substituted into equation (\ref{eq3-23}). The density contrast
$\delta(x,\theta,\phi)$ in real space is also expanded by normal modes
as in equation (\ref{eqa3-15a}). In this way, the integration by
$(x',\theta',\phi')$ in equation (\ref{eq3-23}) is explicitly
performed by orthogonal relations of equations (\ref{eqa2-2-1a}),
(\ref{eqa2-6a}), resulting in
\begin{equation}
  \delta_R(x,\theta,\phi) = 
  \sum_{l,m} \int\frac{k^2dk}{2\pi^2} W(kR)
  \widetilde{\delta}_{lm}(k) X_l(k,x) Y_l^m(\theta,\phi).
\label{eq3-29}
\end{equation}
This equation is a familiar one in the Fourier-Bessel expansion of
smoothed fields in the flat universe, and is a natural extension to
the non-flat universe, in general. The expansion (\ref{eq3-29}) has
just the same form as in the unsmoothed field of equation
(\ref{eqa3-15a}), with the replacement of $\widetilde{\delta}_{lm}(k)
\rightarrow W(kR)\widetilde{\delta}_{lm}(k)$. Consequently, the
similar derivation of the unsmoothed correlation function in redshift
space applies and the result of the smoothed correlation function is
just given by the replacement $P(k) \rightarrow W^2(kR) P(k)$ in the
final expression of the unsmoothed expression in Appendix~\ref{app2}
[eqs. (\ref{eqa3-17})--(\ref{eqa3-20i})]. As a result, the correlation
matrix (\ref{eq3-20}) in redshift space with spherical cells of a
fixed smoothing radius $R$ in comoving space is given by
\begin{equation}
  \xi^{\rm (s)}_{ij} = 
  b_i b_j D_i D_j
  \sum_{n=0}^2 \sum_{l=0}^{2n}
  c_l^{(n)}(x_i,x_j,\theta_{ij}) \xi_l^{(n)}(x_{ij}),
\label{eq3-30}
\end{equation}
where
\begin{equation}
  \xi^{(n)}_l(x) \equiv
  \frac{(-1)^n}{{S_K}^{2n-l}(x)}
  \int\frac{k^2dk}{2\pi^2}
  \frac{\sqrt{N_l(k)}}{k(k^2 - 4K)^n} X_l(k,x) W^2(kR) P(k),
\label{eq3-31}
\end{equation}
the factor $N_l(k)$ is given by equation (\ref{eqa2-5}) and the
functions $c^{(n)}_l(x_i,x_j,\theta_{ij})$ are given by equations
(\ref{eqa3-20a})--(\ref{eqa3-20i}). These equations, i.e., equations
(\ref{eq3-30}), (\ref{eq3-31}), (\ref{eqa3-20a})--(\ref{eqa3-20i}) are
sufficient to construct a correlation matrix. The functions
$c^{(n)}_l$ are given by elementary functions and the numerical
evaluations are very fast. Although the functions $\xi^{(n)}_l$
involve integrations, this one-dimensional function is a common
function to obtain the correlation matrix for a fixed cosmological
model. Once the functions $\xi^{(n)}_l$ are evaluated and tabulated,
one can use appropriate interpolations without integrations cell by
cell. Therefore, one can make the evaluation of $\xi^{(n)}_l(x_{ij})$
a very fast procedure. These form our main technique of the fast
construction of the correlation matrix.

The simplicity of constructing the correlation matrix is largely
dependent of restricting ourselves in linear theory. On nonlinear
scales, the shape of the power spectrum is evolving, and the redshift
distortions are not properly described by predictions of linear
theory. To ensure the nonlinearity does not have significant effects,
one should carefully choose the KL modes which are described by linear
theory. This is achieved by calculating Fourier spectra of KL modes in
indivisual survey \citep[see, e.g.,][]{pop04}.

In practice, the comoving separations $x_{ij}$ are usually not
comparable to the curvature scale of the universe. Observationally,
$|\mOmega_{\rm K0}|$ is less than $0.04$ \citep{ben03,spe03}, which
means the curvature scale is approximately five times larger than the
Hubble scale, $|K|^{-1/2} \simgt 5c/H_0$. The clustering scale we
observe is usually much smaller than the Hubble scale. In this case,
we can safely put $K \rightarrow 0$ in the equation (\ref{eq3-31}),
since $x_{ij} \ll |K|^{-1/2}$. That is,
\begin{equation}
  \xi^{(n)}_l(x) =
  \frac{(-1)^{n+l}}{x^{2n-l}}
  \int\frac{k^2dk}{2\pi^2}
  \frac{j_l(kx)}{k^{2n-l}} W^2(kR) P(k),
\label{eq3-32}
\end{equation}
when $x\ll |K|^{-1/2}$. Even in this approximation, the effect of
curvature on the separation between the observer and the sample, $x_i
= x(z_i)$, are still included through the coefficients $c^{(n)}_l$.
The terms with a factor $|K|{S_K}^2(x_{ij})$ in $c^{(n)}_l$'s can also
be neglected when the above approximation $x_{ij} \ll |K|^{-1/2}$ is
applied.

In the shallow redshift surveys, where $z_i \ll 1$ for all cells, one
can totally neglect the curvature effect not only in $x_{ij}$, but
also in $x_i$. Consequently, the relative distance $x_{ij}$ is simply
given by ${x_{ij}}^2 = {x_i}^2 + {x_j}^2 - 2 x_i x_j \cos\theta_{ij}$.
The distances are just given by the redshifts, $x_i = cz_i/H_0$, which
does not depend on any cosmological parameter when distances are
measured in units of $\himpc$. The quantity $\gamma_{ij}$ defined in
equation (\ref{eqa3-22}) is reduced to $\gamma_{ij} = (x_i - x_j
\cos\theta_{ij})/x_{ij}$ and the quantity $\widetilde{\theta}_{ij}$ in
equation (\ref{eqa3-23}) is simply given by $\widetilde{\theta}_{ij} =
\theta_{ij}$. In the expression of the coefficients $c^{(n)}_l$ of
equations (\ref{eqa3-20a})--(\ref{eqa3-20i}), one can replace $\beta_i
= \beta_j = \beta_0 \equiv \beta(0)$, and the terms proportional to
$|K|{S_K}^2(x_{ij})$ are negligible. The function $\alpha(z)$ in the
equation (\ref{eqa3-16-2}) or (\ref{eqa3-16-3}) is predominantly
determined by the variation of the selection function $\mPhi(z)$ or
$n(z)$ in the sample, so that the factor $A_{ij}$ in the equation
(\ref{eqa3-21}) is given by
\begin{equation}
  A_{ij} = \frac{x_{ij}}{x_i}
  \left\{
    2 + \left.\frac{d\ln \mPhi(z)}{d\ln z}\right|_{z_i}
  \right\} =
  \frac{x_{ij}}{x_i} \left.\frac{d\ln n(z)}{d\ln z}\right|_{z_i}
\label{eq3-33}
\end{equation}
The resulting formula of the correlation function is equivalent to the
expression derived by \citet{sza98}. The simplicity of the shallow
surveys is that the correlation matrix depends polynomially on
$\beta_0 = f(0)/b(0)$, and that there are no other dependences on the
cosmological parameters, when the power spectrum is given.

In the limit $x_{ij} \ll x_i, x_j$, the distant-oberver approximation
is fulfilled, which drastically reduces the complexity of the formula
\citep{mat00}. In this limit, one can reasonably assume that the
quantity $H(z)D(z)f(z)n(z)$ in the equation (\ref{eqa3-16-3}) is
approximately the same at $z_i$ and $z_j$ so that $A_{ij} = 0$ in the
equation (\ref{eqa3-21}). In the same limit, the approximation
$\gamma_{ij} = \pi - \gamma_{ji}$, and $\cos\widetilde{\theta}_{ij} =
1$ are followed. As a result, the coefficients $c^{(n)}_l$ of
equations (\ref{eqa3-20a})--(\ref{eqa3-20i}) survive only for
\begin{eqnarray}
&&
  c^{(0)}_0 = 1 + \frac13\left(\beta_i + \beta_j\right) +
  \frac15 \beta_i \beta_j,
\label{eq3-34a}\\
&&
  c^{(1)}_2 =
  \left[
    \frac23\left(\beta_i + \beta_j\right) +
    \frac47\beta_i\beta_j
  \right] P_2(\cos\gamma),
\label{eq3-34b}\\
&&
  c^{(2)}_4 =
    \frac{8}{35}\beta_i\beta_j P_4(\cos\gamma),
\label{eq3-34c}
\end{eqnarray}
where $P_l$ are the Legendre polynomials, and $\gamma \equiv
\gamma_{ij}$ is the angle between the line of sight and the separation
$x_{ij}$. Although the distant-observer approximation is simpler, it
is not appropriate in wide-angle redshift surveys. The general formula
has more terms than the distant-observer approximation. However, this
does not bring any significant problem to numerical evaluations of the
correlation matrices.

\begin{figure}[ht]
\epsscale{0.95} \plotone{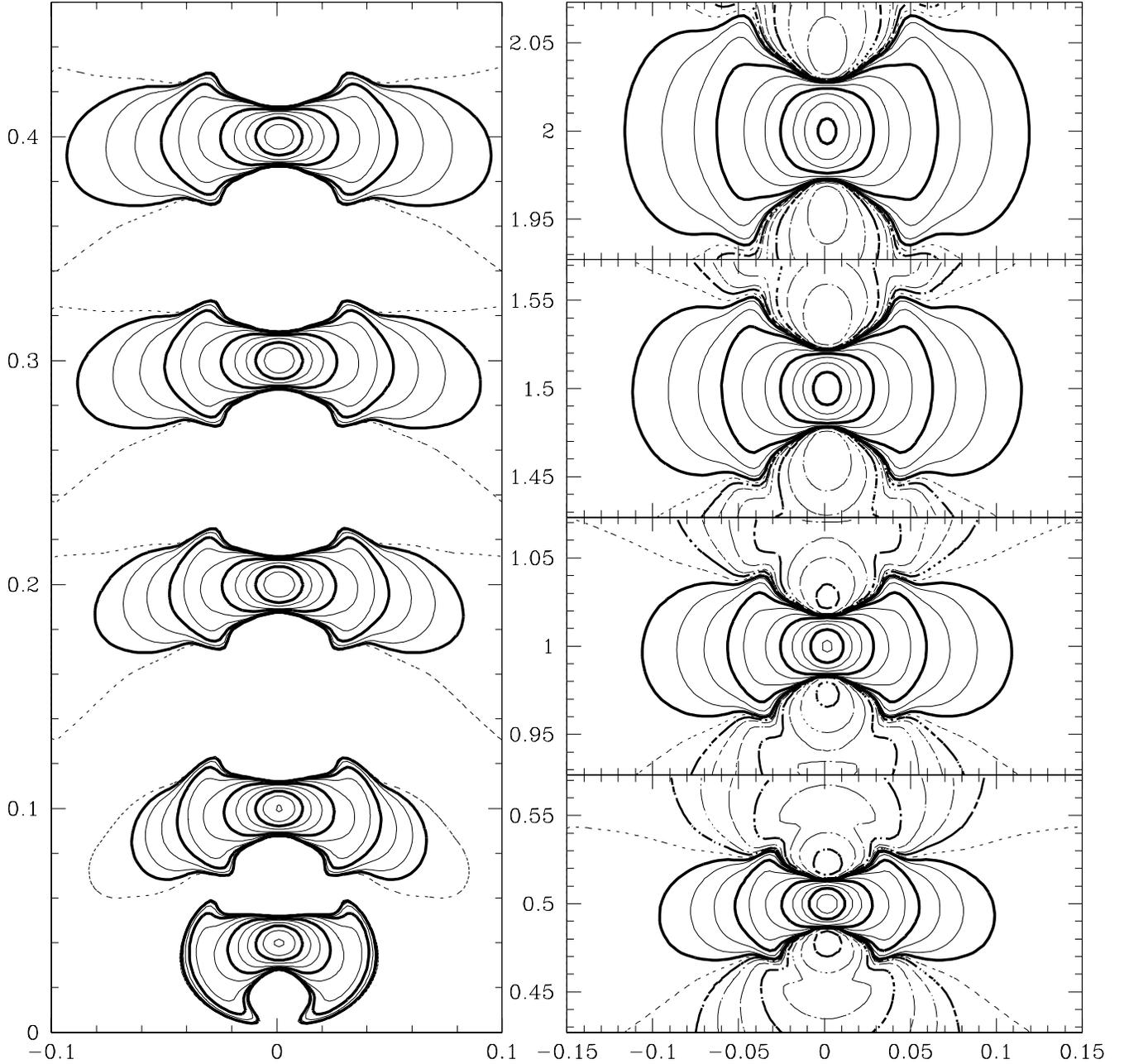} \figcaption[f_xis.eps]{ The
smoothed ($R = 15\himpc$) correlation function in redshift space in
the concordance model: $\mOmega_{\rm M0} = 0.3$, $\mOmega_{\rm K0} =
0$, $w = -1$, $h = 0.7$, $f_{\rm baryon} = 0.15$, $b=1$, $\sigma_8 =
1$. Contour lines indicate the value of the smoothed correlation
function around the centers at redshifts $z_i=0.04, 0.1, 0.2, 0.3,
0.4, 0.5, 1, 1.5, 2$. The coordinate axes correspond to $(z_j
\sin\theta_{ij}, z_j \cos\theta_{ij})$, i.e., the direct $z$-space.
The contours have the intervals of $\Delta \log_{10} \xi_{ij} = 1/3$.
Thick contours indicate $\xi_{ij} = 0.1, 0.01, 10^{-3}$, $10^{-4}$
from inner to outer contours, respectively. The zero-points of
$\xi_{ij}$ are plotted by dotted lines. In the right panels, negative
regions are also shown by dashed contours with the same contour levels
with minus signs.
\label{fig-xis}}
\end{figure}

As an illustration, the resulting smoothed correlation functions
calculated by our methods are plotted in Figure~\ref{fig-xis}. We
assume the cosmological parameters are those of the concordance model,
$\Omega_{\rm M0} = 0.3$, $\Omega_{\rm K0} = 0$, $w = -1$, $h = 0.7$,
$f_{\rm baryon} = 0.15$, $b=1$, $\sigma_8 = 1$, where the Hubble
constant in units of 100km/s/Mpc, $h$, the baryon fraction $f_{\rm
baryon}$, and the normalization of the mass fluctuations $\sigma_8$
are needed only for modeling the CDM + baryon power spectrum $P(k)$
\citep[we used the fitting formula by][]{eis98}. The smoothing radius
is $15\himpc$. In the Figure, the contour lines of the correlation
function around centers at redshifts $z=0.04, 0.1, 0.2, 0.3, 0.4, 0.5,
1, 1.5, 2$ are plotted. The plots are in ``$z$-space'' where the
observer is sitting at the origin (0,0). More precisely, the contour
lines indicates the values of the correlation matrix $\xi^{\rm
(s)}_{ij}$, where $z_i$ takes the above discretized values. The
remaining variables $z_j$ and $\theta_{ij}$ correspond to the
coordinates $(z_j \sin\theta_{ij}, z_j \cos\theta_{ij})$ in the
Figure.

At low redshifts, the deviations from the distant-observer
approximation are apparent. Thoughtless application of the
distant-observer approximation results in a fatal error. Althought one
can still choose parameterization in low-redshift correlations to
obtain better results, our method is free from this ambiguity.

\begin{figure}[ht]
\epsscale{1.0} \plotone{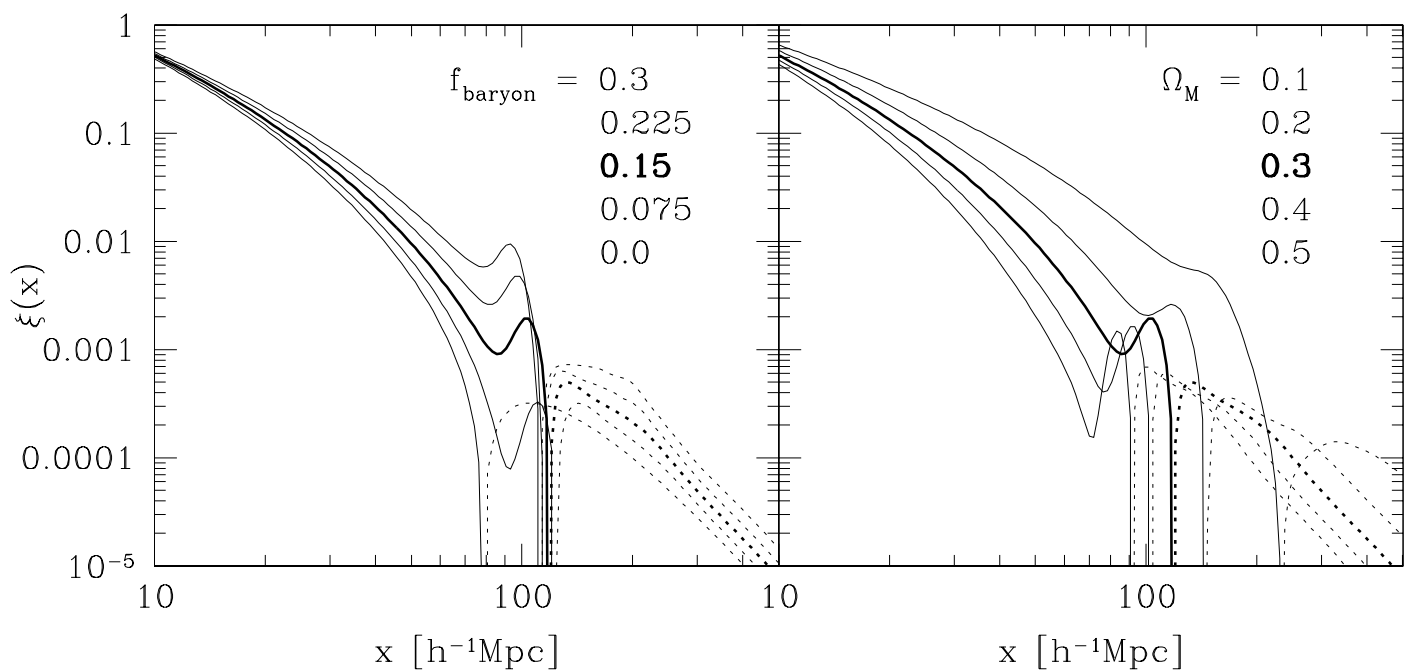} \figcaption[f_xir.eps]{ The
correlation function in real space, $\xi(x)$. The thick lines
correspond to the concordance model which is the same as in the
Figure~\ref{fig-xis}. The baryon fraction $f_{\rm baryon}$ is varied
in the left panel, and the total density parameter $\mOmega_{\rm M0}$
is varied in the right panel, as indicated in the figures. Higher
baryon fraction and lower density parameter both give higher
amplitudes of the correlations. Dotted lines indicate the negative
correlations.
\label{fig-xir}}
\end{figure}

There are prominent bumps in the contours which makes the contours
look peculiar. These bumps are attributed to the baryon oscillation in
the power spectrum. In fact, the real-space correlation function
$\xi(x)$ has the corresponding bump on the scale of around
$100\himpc$, as shown in Figure~\ref{fig-xir}. The thick curves
correspond to the correlation function of the model which is
equivalent to that used in Figure~\ref{fig-xis}. There is a sharp peak
on the scale of $100\himpc$. This peak does not exist when $f_{\rm
baryon}=0$, as also shown in the Figure.

When the smoothing is applied, the sharpness of the peak is reduced,
and this ``baryon peak'' will be more like ``baryon shoulder'' in the
correlation function. This baryon shoulder introduces a scale on the
correlation function, which can be used as a ``standard ruler'' to
measure the geometry of the universe \citep{mat03,bla03,seo03}. The
location of the baryon peak in $\xi(x)$ is sensitive to the parameter
$\mOmega_{\rm M0}h$ which determine the horizon size at the time of
equality. The amplitude of the peak is sensitive to the parameter
$f_{\rm baryon}$ which determine the strength of the baryon
oscillation. Because the geometry of the universe determines the
apparent scale of the baryon shoulder, the observation of $\xi_{ij}$
constrains the geometry, almost irrespectively of the value of $f_{\rm
baryon}$ \citep{mat02}.


\subsection{
Spherical Approximations of the Kernel
}

In the construction of the correlation matrix in the previous
subsection, we have assumed the sphericity of the kernel in comoving
space. In reality, the comoving distance--redshift relation and the
curvature of the universe is needed in order to define a cell in
redshift space, that has spherical comoving shape. This means that we
have to assume the parameters $\mOmega_{\rm M0}$, $\mOmega_{\rm Q0}$,
and $w(z)$ to exactly set the spherical shape. If we assume a wrong
geometry, the assumed comoving space is elongated or squashed to the
line-of-sight direction. As a result, we end up with spherioidal cells
in comoving space. One of the important application of calculating the
correlation matrix in this paper is to determine cosmological
parameters by Bayesian analysis, so that the uncertainty in setting
spherical cells above is not desirable. However, the calculation of
the correlation matrix for spheroidal cells is more complicated than
spherical cells. The higher-order multipoles of the spheroidal cells
cause an additional convolution in the expression (\ref{eq3-29}). The
simplicity of the correlation matrix is not preserved in that case,
and thus the numerical computation of the correlation matrix will be
significantly slowed down.

However, such elongation or squashing does not significantly change
the value of the correlation matrix. The elements of the correlation
matrix for pairs of distant cells are virtually unchanged, since the
correlation function is a smooth function and the elements of the
correlation matrix for distant cells are virtually the same as a bare
values of the correlation function without smoothing. The elements of
the correlation matrix for nearby cells are affected by a shape of the
cells, although the effect is not expected to be large as long as the
elongation or squashing is not large.

Keeping the simplicity of the expression of the correlation matrix, we
introduce an approximate method to calculate the correlation matrix
using spherical cells even when the assumed geometry is not exactly
true. In this method, first we adopt a best guess of the geometry with
assumed set of parameters, $\mOmega_{\rm M0}$, $\mOmega_{\rm Q0}$, and
$w(z)$. The KL-mode projection in the Bayesian analysis explained in
the previous section also needs a fiducial cosmological model, so that
it might be natural to take the same parameters as for that fiducial
model, although it is not necessary. Next we set the cells in observed
redshift space to be spherical in the comoving space of the assumed
geometry, and construct a data vector by galaxy counts in these cells.
The assumed model to construct the data vector will be called the
``map model'' below. The cell radius $R$ is defined in this map model.

In the Bayesian analysis, the data vector is fixed throughout the
parameter estimation where cosmological parameters are varied. The
parameters in the map model are generally not the same as the various
cosmological parameters to be tested in the analysis. Therefore, the
shape of the cells in the constructed data vector is not exactly
spherical in each cosmological model to be tested, and is generically
distorted to spheroids. In our approximation, the spheroidal shape in
each cosmological model is approximated to the spherical shape with
the same volume as the actual spheroids. The volume of the spheroids
are given by
\begin{equation}
  V_{\rm spheroids}(z) =
  \left[\frac{D_{\rm A}(z)}{D_{\rm A}^{\rm(map)}(z)}\right]^2
  \frac{H^{\rm(map)}(z)}{H(z)}
  \frac{4\pi R^3}{3},
\label{eq3-40}
\end{equation}
where $H^{\rm (map)}(z)$ and $D_A^{\rm (map)}(z)$ are the Hubble
parameter and the comoving angular diameter distance of the map model,
and $H(z)$ and $D_A(z)$ are the counterparts of the tested model in
the parameter estimation. The spheroidal cell is approximated to the
spherical cell of the effective radius,
\begin{equation}
  R_{\rm eff}(z) = 
  \left[\frac{D_{\rm A}(z)}{D_{\rm A}^{\rm(map)}(z)}\right]^{2/3}
  \left[\frac{H^{\rm(map)}(z)}{H(z)}\right]^{1/3} R.
\label{eq3-41}
\end{equation}

Since this kind of the shape correction is necessary only for nearby
cells, and the effective radius is almost the same for nearby cells,
we do not need to consider the smoothing integration of the cross
correlation with different smoothing radius, in practice.
Interpolation of the correlation matrices with several effective
smoothing radii is sufficient when the map model and the tested model
is not too different. In defining the map model, it is natural to use
the same cosmological parameters as in constructing the projection
matrix $P$. Therefore, when the analysis is iterated with respect to
optimizing the projection matrix, it is preferable that the
cosmological parameters of the map model are also iterated, even
though it is not strictly necessary.


\subsection{
Insignificance of the Finger-of-God Effects on Linear Scales }

While the linear velocity field induces only the coherent motion of
galaxies, random velocities are also present in the nonlinear regime,
and the clustering shape in redshift space is elongated in the
direction of line of sight. This is the famous finger-of-God effect
which is not described by the linear formula. How does the
finger-of-God affect on the correlation matrix? Since the
finger-of-God effect decreases with distances between centers of the
cells, the maximal effect is on the identical cells, that is, on the
variance $\sigma^2(R,z) = \xi^{\rm (s)}_{ii}$. Consequently, we can
estimate the maximal influence of the finger-of-God effect on the
correlation matrix by the same effect on the variance.

The finger-of-God effect can be analytically modeled by incoherent
peculiar velocities on small scales \citep{pea92,pea94}. We can safely
use the distant-observer approximation in this case. When the
incoherent motion along the line of sight is empirically fitted by an
exponential distribution function, $f(v) =
\exp(-\sqrt{2}|v|/\sigma_{\rm p})/(\sqrt{2} \sigma_{\rm p})$
\citep{dav83}, the power spectrum with the finger-of-God effect is
given by \citep{par94}
\begin{equation}
  P_{\rm FOG}(k,\mu) = 
  \frac{P(k)}{\left(1 + k^2 \mu^2 {\sigma_{\rm F}}^2 /2\right)^2},
\label{eq3-50}
\end{equation}
where $\mu$ is the direction cosine of the wavevector $\bfm{k}$ with
respect to the line of sight, and $\sigma_{\rm F} = \sigma_{\rm
p}/H(z)$ is the rms of the displacement in redshift space by
incoherent velocities. The element of the correlation matrix,
including linear redshift-distortion, is therefore modeled by
\begin{equation}
  \xi_{ij} =
  \int \frac{d^3k}{(2\pi)^3} e^{i\sbfm{k}\cdot\sbfm{x}_{ij}}
  W^2(kR)
  \left(
    \frac{1 + \beta \mu^2}
      {1 + k^2 \mu^2 {\sigma_{\rm F}}^2 /2}
  \right)^2
  P(k),
\label{eq3-51}
\end{equation}
where $\bfm{x}_{ij} = \bfm{x}_i - \bfm{x}_j$, $\mu = k_\parallel/k$
and $k_\parallel$ is the line-of-sight component of the wavevector
$\bfm{k}$. Adopting the polar coordinates $\bfm{k} = k (\sin\theta
\cos\phi, \sin\theta \sin\phi, \cos\theta)$ in $k$-space, and setting
$\bfm{x}_{ij} = (x_\perp, 0, x_\parallel)$, the $\phi$-integration can
be analytically performed to give
\begin{equation}
  \xi_{ij} =
  \int \frac{k^2dk}{2\pi^2}  W^2(kR)  P(k)
  \int_{-1}^1 \frac{d\mu}{2}
  \cos\left(k x_\parallel \mu\right)
  J_0\left(k x_\perp \sqrt{1 - \mu^2}\right)
  \left(
    \frac{1 + \beta \mu^2}
      {1 + k^2 \mu^2 {\sigma_{\rm F}}^2 /2}
  \right)^2.
\label{eq3-52}
\end{equation}
The dimensionality of the above integration will not be analytically
reduced further for arbitrary separation $\bfm{x}_{ij}$. The integrand
is a strongly oscillating function, so that the two-dimensional
integration will be a slow process.

However, the correction of the finger-of-God effect is not nessesarily
needed as we will see below. The finger-of-God effect erases the
clustering on scales smaller than $\sigma_{\rm F}$. Observationally,
the velocity dispersion is approximately given by the figure
$\sigma_{\rm p} = 340 {\rm km/s}$ \citep{dav83} for $z=0$. This
dispersion corresponds to $\sigma_{\rm F} = 3.4 \himpc$, which is
smaller than the nonlinearity scale, $R\sim 10\himpc$ for $z=0$.
Therefore, the finger-of-God effect is expected to be smaller than the
nonlinear effect on the clustering on average. Higher redshift regions
have lower $\sigma_{\rm p}$ so that the finger-of-God effect is less
significant.

To find more quantitative estimates of what is the finger-of-God
effect on the correlation matrix, we compare the variance with and
without the finger-of-God effect. The non-diagonal elements of the
correlation matrix are less affected. The variance $\sigma^2(R)$ is
given by substituting $x_\perp = x_\parallel = 0$ in the equation
(\ref{eq3-52}), and the $\mu$-integration can be analytically
performed to give \citep{col95}
\begin{eqnarray}
&&
  \sigma^2(R) =
  \frac12 \int \frac{k^2dk}{2\pi^2}
  \left\{
    \frac{1}{1 + k^2 {\sigma_{\rm F}}^2/2} +
    \frac{{\rm Arctan}\left(k\sigma_{\rm F}/\sqrt{2}\right)}
      {k \sigma_{\rm F}/\sqrt{2}}
  \right.
\nonumber\\
&& 
  \left.\qquad\qquad\qquad -\,
    \frac{2}{k^2 {\sigma_{\rm F}}^2/2}
    \left[
      \frac{1}{1 + k^2 {\sigma_{\rm F}}^2/2} -
      \frac{{\rm Arctan}\left(k\sigma_{\rm F}/\sqrt{2}\right)}
        {k \sigma_{\rm F}/\sqrt{2}}
    \right] \beta
  \right.
\nonumber\\
&& 
  \left.\qquad\qquad\qquad +\,
    \frac{3}{k^4 {\sigma_{\rm F}}^4/4}
    \left[
      \frac{1 + k^2 {\sigma_{\rm F}}^2/3}{1 + k^2 {\sigma_{\rm F}}^2/2} -
      \frac{{\rm Arctan}\left(k\sigma_{\rm F}/\sqrt{2}\right)}
        {k \sigma_{\rm F}/\sqrt{2}}
    \right] \beta^2
  \right\}
  W^2(kR) P(k).
\label{eq3-53}
\end{eqnarray}
Taking the limit $\sigma_{\rm F} \rightarrow 0$, a familar
amplification factor $1 + 3\beta/2 + \beta^2/5$ appears \citep{kai87}.
Focusing only on the finger-of-God effect, we set $\beta = 0$ for the
moment. As an illustration, the variance in real space and in redshift
space (with only the finger-of-God effect) are shown in
Table~\ref{tab1}.
\begin{table}
\begin{center}
\caption{The finger-of-God effect on the variance for the power-law
  model with $n = -1.2$.
\label{tab1}}
\begin{tabular}{c|cccccc}
\tableline\tableline
$R (\himpc)$ & 5 & 10 & 15 & 20 & 30 & 50\\
\tableline
$\sigma_{\rm real}^2(R)$ & 2.3 & 0.67 & 0.32 & 0.19 & $9.3 \times
10^{-2}$ & $3.69 \times 10^{-2}$\\
$\sigma_{\rm FOG}^2(R)$ & 1.7 & 0.59 & 0.30 & 0.18 & $9.1 \times
10^{-2}$ & $3.66 \times 10^{-2}$\\
difference & 28\% & 12\% & 6.4\% & 4.0\% & 2.0\% & 0.8\%\\
\tableline
\end{tabular}
\end{center}
\end{table}
We take a power-law spectrum $P(k) \propto k^{-1.2}$ which corresponds
to observations in nonlinear regime. The normalization
$\sigma(R=8\himpc) = 1$, and the velocity dispersion $\sigma_{\rm p} =
340 {\rm km/s}$ are adopted, which match the observations
\citep{dav83}. The power spectrum on large scales deviates from the
power-law spectrum. In Table~\ref{tab2} the results for the cold dark
matter (CDM) spectrum with a fitting formula of \citet{eis98} are
shown.
\begin{table}
\begin{center}
\caption{The finger-of-God effect on the variance for the CDM model
  with $\mGamma = 0.2$.
\label{tab2}}
\begin{tabular}{c|cccccc}
\tableline\tableline
$R (\himpc)$ & 5 & 10 & 15 & 20 & 30 & 50\\
\tableline
$\sigma_{\rm real}^2(R)$ & 1.9 & 0.71 & 0.36 & 0.22 & $8.4 \times
10^{-2}$ & $3.50 \times 10^{-2}$\\
$\sigma_{\rm FOG}^2(R)$ & 1.5 & 0.64 & 0.34 & 0.21 & $8.3 \times
10^{-2}$ & $3.46 \times 10^{-2}$\\
difference & 22\% & 10\% & 5.3\% & 4.8\% & 1.5\% & 1.2\%\\
\tableline
\end{tabular}
\end{center}
\end{table}
The damping by the finger-of-God does not dominate the linear
infall effect on larger scales. The linear infall effect
enhances the variance by a factor $1 + 3\beta/2 + \beta^2/5$. When
$\mOmega_{\rm M0} = 0.3$ and $b = 1$, the enhancement factor is
$37\%$, and when $b=2$, the factor is $17\%$. These figures are always
larger than the finger-of-God effect when $R \simgt 10 \himpc$.

The dynamical nonlinear effect on the variance is actually larger than
the finger-of-God effect. In fact, the nonlinear correction to the
variance with smoothing is of order ${\sigma_{\rm L}}^2$, where
${\sigma_{\rm L}}^2$ is the variance of the linear theory. For
example, the nonlinear loop correction to the variance without
smoothing is given by $\langle\delta^2\rangle = {\sigma_{\rm L}}^2 +
1.82 {\sigma_{\rm L}}^4$ \citep{sco96}. From Table~\ref{tab1}, we see
that the order of nonlinear corrections are always larger than the
order of finger-of-God corrections by about a factor of five. Even
when the radius of the cell $R$ is still on nonlinear scales, the
KL-mode projection can only leave the linear regime. In which case,
the nonlinear correction is not needed in the evaluation of the
correlation matrix, neither is the finger-of-God effect.


\subsection{
Correlation Matrices in Projected Samples on the Sky
}

When only the position on the sky is catalogued and the redshifts of
galaxies are not observed, the data vector is a set of galaxy counts
on the 2-dimensional sky, projected along the line of sight with some
redshift distribution function $n(z)$. When the redshifts are totally
unobserved, the selection function spreads over a wide redshift range.
With photometric redshift data, the selection function can be chosen
so that the selection is approximately uniform in some narrower
redshift range and rapidly drops outside that range.

One of the advantage of the projected sample is the absence of
redshift-distortions. The correlation matrix is given by just
integrating the angular correlation function $w(\theta)$. Handling the
correlation matrix is thus simpler than the 3D sample. When the 2D
smoothing kernel is circularly symmetric, the elements of the
correlation matrix are represented by just a function of the distance
between centers of cells. Therefore, the advantage of the circular
cells is obvious. We consider an efficient method to obtain 2D
correlation matrices for the circularly symmetric cells below.

In projected samples on the sky, the angular correlation function
$w(\theta)$ is given by
\begin{equation}
  w(\theta) =
  \int_0^\infty dz_1 dz_2
  n(z_1) n(z_2) D(z_1) D(z_2) b(z_1) b(z_2)
  \xi[x(z_1,z_2,\theta)],
\label{eq3-60}
\end{equation}
where $x(z_1,z_2,\theta)$ is the comoving distance between two points
with redshifts $z_1$ and $z_2$, and with an opening angle $\theta$.
This function is given by a geometric relation,
\begin{eqnarray}
&&
  {S_K}^2[x(z_1,z_2,\theta)] = 
  {S_K}^2[x(z_1)] + {S_K}^2[x(z_2)]
  - 2 C_K[x(z_1)] C_K[x(z_2)] S_K[x(z_1)] S_K[x(z_2)] \cos\theta
\nonumber\\
&&\qquad\qquad\qquad\qquad
  - K {S_K}^2[x(z_1)] {S_K}^2[x(z_2)] \left(1 + \cos^2\theta\right),
\label{eq3-60-1}
\end{eqnarray}
where $x(z)$ is the comoving distance-redshift relation of equation
(\ref{eq3-4}). The selection function $n(z)$ is the normalized
redshift distribution $dN/dz$ of galaxies with
\begin{equation}
  \int_0^\infty dz\; n(z) = 1.
\label{eq3-61}
\end{equation}
The volume factor is included in $n(z)$ so that the relation to the
selection function per comoving volume $\mPhi(z)$ is given by $H(z)
n(z) \propto {S_K}^2[x(z)] \mPhi(z)$. The correlation function in real
space $\xi(x)$ is given by
\begin{equation}
  \xi(x) = \int \frac{k^2 dk}{2\pi^2}
  \frac{\sin kx}{k S_K(x)} P(k),
\label{eq3-62}
\end{equation}
which is derived from equations (\ref{eqa3-15a}), (\ref{eq3-27}) and
(\ref{eq3-28}) [the equivalent formula of $\xi(x)$ for an open
universe is given in \citet{wil83}].

\begin{figure}[ht]
\epsscale{0.8} \plotone{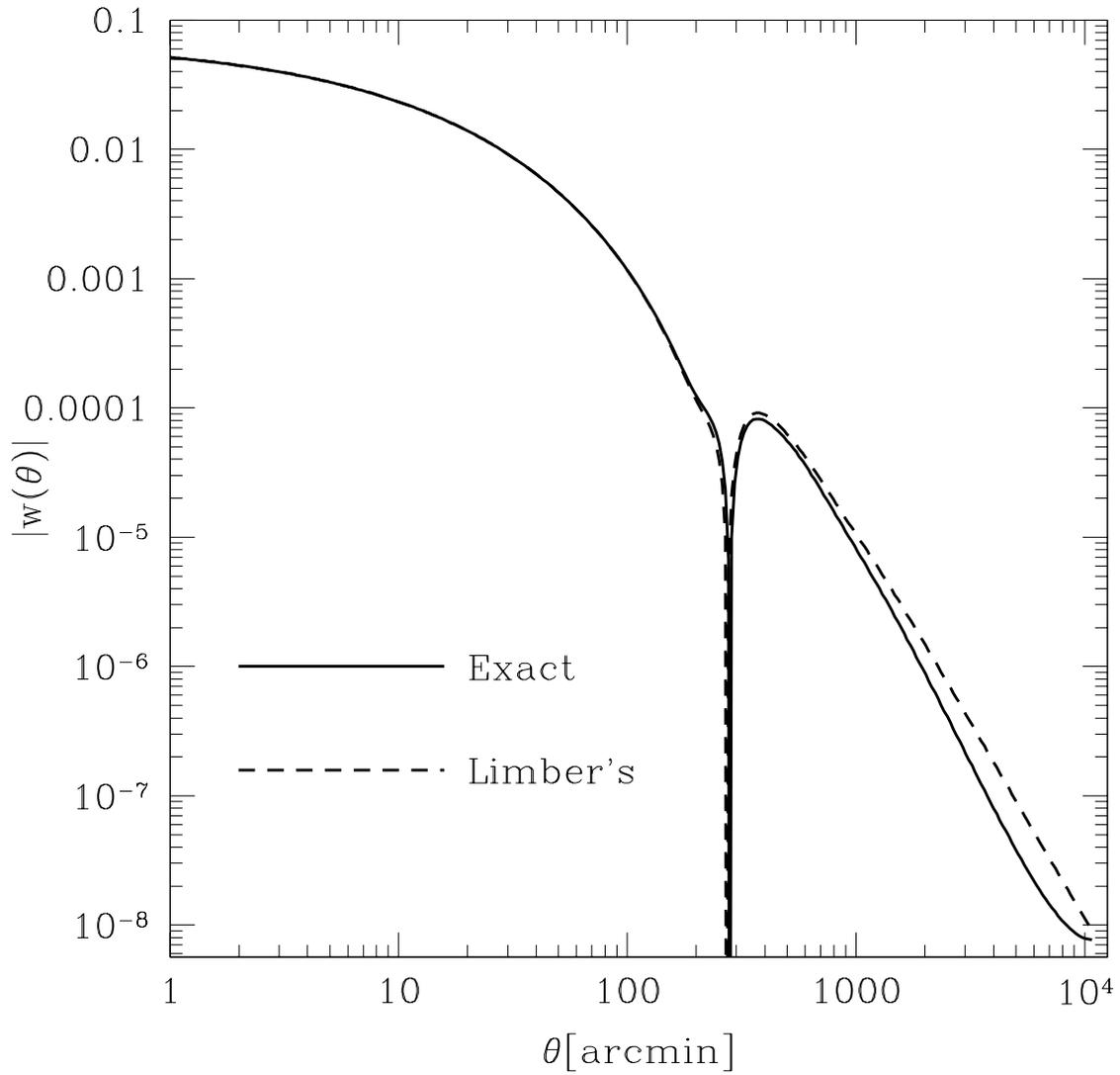} \figcaption[f_wth.eps]{ Angular
correlation function for the concordance model. The solid line shows
the exact prediction of the linear theory (see text). The dashed line
shows the prediction from Limber's equation.
\label{fig-wth}}
\end{figure}

The direct 2-dimensional integration of equation (\ref{eq3-60}) is
numerically straightforward and relatively fast, once the
3-dimensional correlation function $\xi(x)$ is tabulated and
interpolated. In this integration, we should use a dense sampling for
small separation, $x(z_1,z_2,\theta)$, for a fast calculation. For
example, the integration variable $z_2$ for a fixed $z_1$ can be
transformed to $y_{\pm} = |z_2 - z_1| + 2z_1\sin(\theta/2)$. According
to the sign of $z_2 - z_1$ we use $y_+$ and $y_-$ respectively.
Logarithmically uniform sampling in $y_{\pm}$-integration can achieve
the dense sampling for small separations. In this paper, we follow
this strategy in numerically evaluating the equation (\ref{eq3-60}).
The result of the numerical integration for a concordance model,
$\mOmega_{\rm M0} = 0.3$, $\mOmega_{\rm K0} = 0$, $w = -1$, $h = 0.7$,
$f_{\rm baryon} = 0.15$, $b=1$, $\sigma_8 = 1$ is shown on
Figure~\ref{fig-wth} by a solid line. The selection function is simply
taken as a step-wise function, $n(z) = {\rm const.}$ for $0.5 < z < 1$
and $n(z) = 0$ otherwise. In the angular correlation function, the
baryon peak in $\xi(x)$ is almost erased, although there is a hint of
a bump around 100--200 arcmin.

The signal part of the correlation matrix is given by the smoothing
integration of $w(\theta)$:
\begin{equation}
  w_{ij} = 
  \int \sin\theta d\theta d\phi
  \int \sin\theta' d\theta' d\phi'
  W_i(\theta,\phi)
  W_j(\theta',\phi')
  w(\theta_{\rm rel}),
\label{eq3-63}
\end{equation}
where $W_i(\theta,\phi)$ is the smoothing kernel of a cell $i$ with a
unit normalization, and $\theta_{\rm rel}$ is the angle between
directions $(\theta,\phi)$ and $(\theta',\phi')$. When the smoothing
kernel does not have simple shape, or varies from cell to cell, the
straightforward evaluation of the correlation matrix is given by
directly integrating equations (\ref{eq3-60}) and (\ref{eq3-63}).
Firstly, the correlation function $\xi(x)$ is calculated and
tabulated. Secondly, using the interpolation of the tabulated
$\xi(x)$, the two-dimensional integration of equation (\ref{eq3-60})
yields the angular correlation function $w(\theta)$ which is again
tabulated. Finally, the four-dimensional integration of equation
(\ref{eq3-63}) provides the correlation matrix.

However, the above procedure is not a fast process because the
smoothing integration involves a four-dimensional integration cell by
cell. In the philosophy that the numerical computation of the
correlation matrix should be maximally reduced, it is better to choose
the shape of the cells as in the three dimensional analysis of the
redshift surveys. The simplest choice is the circularly symmetric
kernel $W_{\theta_{\rm s}}(\theta)$ where $\theta$ is the relative
angle from a center of a kernel, $\theta_{\rm s}$ is the angular
smoothing scale. The normalization of this kernel is given by
\begin{equation}
  2\pi \int_0^\infty \sin\theta d\theta W_{\theta_{\rm s}}(\theta) = 1
\label{eq3-64}
\end{equation}
In this case, the elements of the correlation matrix $w_{ij}$ depend
only on the angular separation of the cells, $\theta_{ij}$, i.e.,
$w_{ij} = u(\theta_{ij})$. Once the function $u(\theta)$ is calculated
and tabulated, the correlation matrix is obtained by simple
interpolations, which are very fast procedures.

The straightforward representation of the function $C(\theta)$ is
provided by the standard multipole expansion:
\begin{equation}
  u(\theta) = \sum_{l=0}^\infty \frac{2l+1}{4\pi} P_l(\cos\theta)
  \left|W_l\right|^2 C_l,
\label{eq3-65}
\end{equation}
where $C_l$ is the angular power spectrum and $W_l$ is the multipole
of the circular window function
\begin{equation}
  W_l = 2\pi 
  \int_0^\pi \sin\theta d\theta W_{\theta_{\rm s}}(\theta)
  P_l(\cos\theta),
\label{eq3-66}
\end{equation}
The angular power spectrum is related to the spatial power spectrum
$P(k)$ by
\begin{equation}
  C_l = 
  4\pi \int_0^\infty \frac{k^2 dk}{2\pi^2}
  \left|\mPsi_l(k)\right|^2 P(k),
\label{eq3-66-1}
\end{equation}
where $\mPsi_l(k)$ is defined by
\begin{equation}
  \mPsi_l(k) =
  \int_0^\infty dz\; n(z) D(z) b(z) X_l[k,x(z)],
\label{eq3-67}
\end{equation}
and $x(z)$ is given by equation (\ref{eq3-4}). Numerical evaluation of
this expression is not so fast, because integrations over $k$ and $z$
and sum over $l$ are involved for oscillating functions, $X_l$ and
$P_l$. In addition, numerical evaluations of $X_l(k,x)$ might be
tricky for a non-flat universe. Thus, it is better to seek a more
efficient method to numerically evaluate the function $u(\theta)$.

If the angle $\theta$ is sufficiently small and the flat-sky
approximation can be applied, the angular correlation function is
given by Limber's equation \citep{pee80}
\begin{equation}
  w(\theta) = \int_0^\infty dz H(z) n^2(z) D^2(z) b^2(z)
  \int_0^\infty \frac{k dk}{2\pi} J_0\left[k\theta D_{\rm A}(z)\right] P(k),
\label{eq3-68}
\end{equation}
where $D_{\rm A}(z) = S_K[x(z)]$ is the comoving angular diameter
distance, and $J_\nu(x)$ is the Bessel function. This approximation is
compared to the exact integration of equation (\ref{eq3-60}) in
Figure~\ref{fig-wth}. As one can see, Limber's equation is reasonably
accurate for $\theta \simlt 100\ {\rm arcmin.} \sim 1.7\ {\rm
degrees}$ within $0.5\%$ for the concordance model.

The two-dimensional Fourier transform (in the flat-sky approximation)
of the angular correlation function is given by
\begin{equation}
  \tilde{w}(l) =
  \int d^2\theta e^{-i\sbfm{l}\cdot\sbfm{\theta}} w(\theta) =
  2\pi \int_0^\infty \theta d\theta J_0(l\theta) w(\theta) =
  \int_0^\infty dz \frac{H(z) n^2(z) D^2(z) b^2(z)}{{d_{\rm cA}}^2(z)}
  P\left(\frac{l}{d_{\rm cA}(z)}\right),
\label{eq3-69}
\end{equation}
where the completeness relation of the Bessel function
\citep[e.g.,][]{arf01},
\begin{equation}
  \int_0^\infty x dx J_\nu(ax) J_\nu(bx) = \frac{1}{a} \delta(a-b)
  \qquad \left(\nu > -1/2 \right)
\label{eq3-70}
\end{equation}
is used. The two-dimensional Fourier window function is given by
\begin{equation}
  W(l\theta_{\rm s}) =
  \int d^2\theta e^{-i\sbfm{l}\cdot\sbfm{\theta}}
  W_{\theta_{\rm s}}(\theta) =
  2\pi \int_0^\infty \theta d\theta J_0(l\theta)
  W_{\theta_{\rm s}}(\theta),
\label{eq3-71}
\end{equation}
which should not be confused with the three-dimensional one in the
previous subsection although we use the same notation. It is
straightforward to show that $\widetilde{w}(l) \simeq C_l$ and
$W(l\theta_{\rm s}) \simeq W_l$ in the limit of $l \gg 1$ because
$P_l (\cos\theta) \simeq J_0(l\theta)$ in the same limit. The top-hat
window function corresponds to $W(l\theta_{\rm s}) = 2
J_1(l\theta_{\rm s})/(l \theta_{\rm s})$. The smoothed angular
correlation function $u(\theta)$ is the convolution of the angular
correlation $w(\theta)$ with the smoothing kernel $W_{\theta_{\rm
s}}(\theta)$ for both ends, and consequently, using the convolution
theorem, we have
\begin{eqnarray}
  u(\theta) &=& \int_0^\infty \frac{l dl}{2\pi}
  J_0(l\theta) W^2(l\theta_{\rm s})\tilde{w}(l)
\nonumber\\
 \qquad\qquad &=&
  \int_0^\infty dz H(z) n^2(z) D^2(z) b^2(z)
  \int_0^\infty \frac{k dk}{2\pi}
  J_0\left[k\theta d_{\rm cA}(z)\right]
  W^2\left[k\theta_{\rm s} d_{\rm cA}(z)\right] P(k).
\label{eq3-72}
\end{eqnarray}
The numerical performance of this two-dimensional integration is
reasonably fast because the integrand of the variable $z$ is a smooth
function. The integration by the variable $k$ is fast as long as the
effective spectral index of $P(k)$ is negative on the relevant scales
which is indicated by the angle $\theta$.

Limber's equation is not appropriate for wide-angle correlations.
However, the smoothing angle $\theta_{\rm s}$ is much smaller than the
opening angle of the survey. In this case, we can apply flat
coordinates within the smoothing cells, and the function $u(\theta)$
is straightforwardly represented by
\begin{eqnarray}
&&
  u(\theta) =
  \int \theta_1 d\theta_1 d\phi_1
  \int \theta_2 d\theta_2 d\phi_2\;
  W_{\theta_{\rm s}}(\theta_1) W_{\theta_{\rm s}}(\theta_2)
\nonumber\\
&& \qquad\qquad\qquad\times\,
  w\left[
    \sqrt{(\theta + \theta_2 \cos\phi_2 - \theta_1 \cos\phi_1)^2 +
      (\theta_2 \sin\phi_2 - \theta_1 \sin\phi_1)^2}
   \right].
\label{eq3-73}
\end{eqnarray}
This four-dimensional integration is not needed for a separation angle
larger than the size of the cells, $\theta \gg \theta_{\rm s}$,
because the smoothing effect is small in that case, so that $u(\theta)
= w(\theta)$ in the lowest (0th) order. Taylor expansion of the
equation (\ref{eq3-73}) with respect to $\theta_1/\theta$ and
$\theta_2/\theta$ yields higher-order corrections to this simple
approximation of the 0th order. The smoothing kernel of the form,
\begin{equation}
  W_{\theta_{\rm s}}(\theta) =
 {\theta_{\rm s}}^{-2} W_{\rm s}(\theta/\theta_{\rm s}),
\label{eq3-74}
\end{equation}
has a scaling with respect to the smoothing angle, where $W_{\rm s}$
is a single-variable function, independent on smoothing angle
$\theta_{\rm s}$, and normalized by
\begin{equation}
 2\pi \int_0^\infty t\, dt\; W_{\rm s}(t) = 1.
\label{eq3-75}
\end{equation}
Defining scaled moments of the kernel,
\begin{equation}
  m_k \equiv 2\pi \int_0^\infty t\,dt\;t^k\;W_{\rm s}(t),
\label{eq3-76}
\end{equation}
the Taylor expansion of the equation (\ref{eq3-73}) up to forth order
is straightforwardly calculated to give
\begin{eqnarray}
&&
  u(\theta) = w(\theta) +
  \frac{m_2}{2}
  \left[w''(\theta) + \frac{w'(\theta)}{\theta}\right]
  {\theta_{\rm s}}^2
\nonumber\\
&&\qquad\qquad
  +\; \frac{m_4 + 2 {m_2}^2}{32}
  \left[
    w''''(\theta) + \frac{2w'''(\theta)}{\theta} -
    \frac{w''(\theta)}{\theta^2} + \frac{w'(\theta)}{\theta^3}
  \right] {\theta_{\rm s}}^4 + \cdots.
\label{eq3-77}
\end{eqnarray}
For a top-hat kernel, $W_{\theta_{\rm s}}(\theta) =
\mTheta(\theta_{\rm s} - \theta)/\pi {\theta_{\rm s}}^2$,
$W_{\rm s}(t) = \mTheta(1-t)/\pi$ and $m_k = 1/(k/2 + 1)$ so that the
numerical coefficients of second-order and forth-order terms are $1/4$
and $5/192$, respectively.

\begin{figure}[ht]
\epsscale{0.8} \plotone{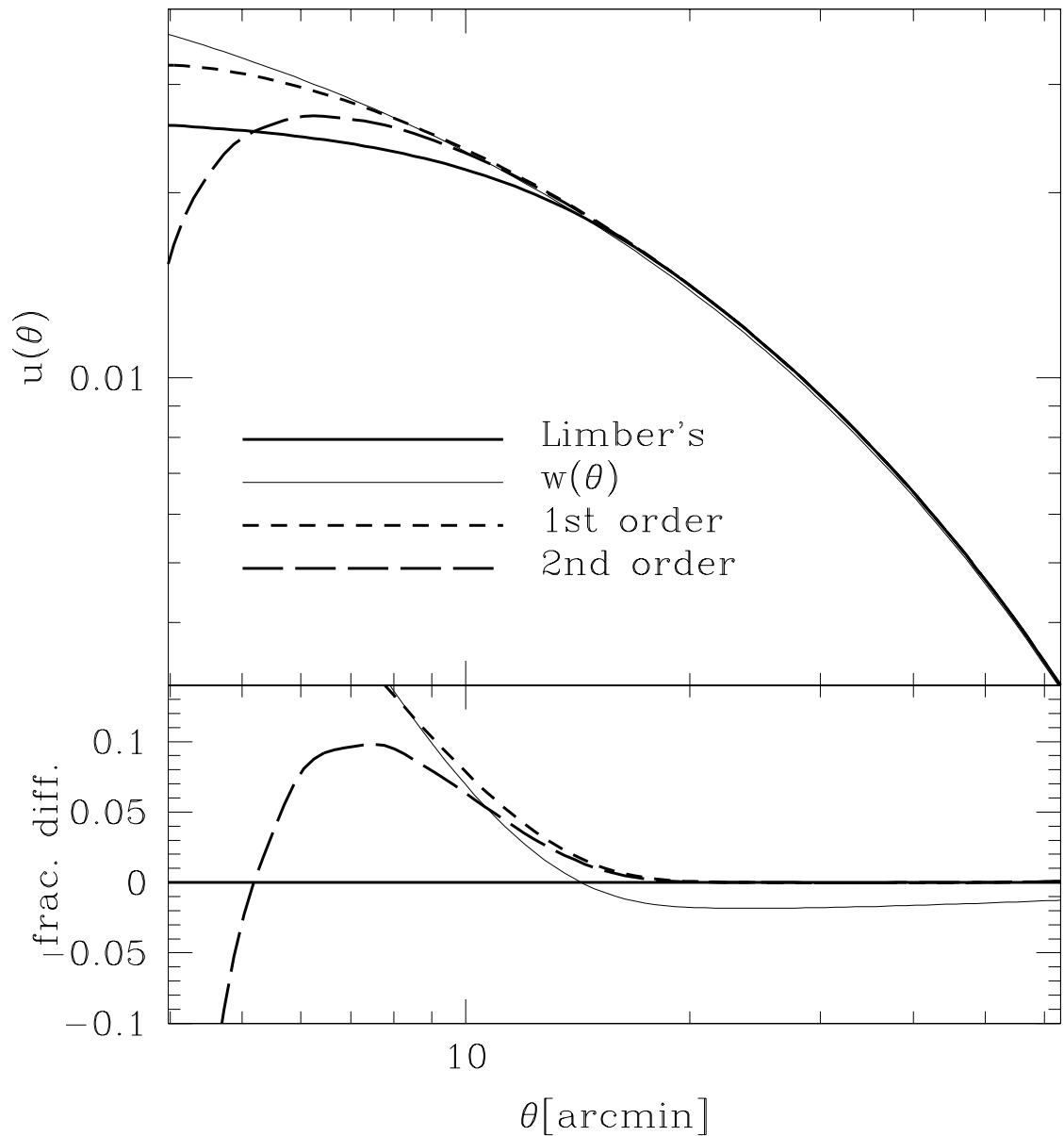} \figcaption[f_ws.eps]{ The smoothed
angular correlation function $u(\theta)$ around the smoothing scale
($\theta_{\rm s} = 10$ arcmin.). Thin solid line: the bare angular
correlations. Thick solid line: smoothed correlations with Limber's
equation. Short-dashed line: Taylor approximation up to 1st order.
Long-dashed line: Taylor approximation up to 2nd order. The lower
panel shows the fractional differences with respect to the prediction
of Limber's equation, which gives the presice value of $u(\theta)$ in
the angle range here.
\label{fig-ws}}
\end{figure}

On Figure~\ref{fig-ws}, the smoothed angular correlation functions
$u(\theta)$ in the various approximations given above are compared
when a top-hat kernel is adopted with a smoothing angle $\theta_{\rm
s} = 10$ arcmin. In the range of angles plotted here, the Limber's
equation is accurate enough and the equation (\ref{eq3-72}) gives
practically exact result. The first-order approximation, in which the
terms up to ${\theta_{\rm s}}^2$ in equation (\ref{eq3-77}) are kept,
well reproduce the exact result for $\theta \simgt 1.6 \theta_{\rm
s}$. The second-order approximation does not significantly improve the
first-order approximation. At $\theta = 2\theta_{\rm s}$, fractional
errors are 4\%, 0.9\%, and 0.3\% for truncations of the Taylor series
up to zeroth, first, and second orders, respectively. Therefore, one
does not need to perform the four-dimensional numerical integration
even for neighboring cells, as long as cells do not overlap. We only
need to numerically evaluate the bare angular correlation function
$w(\theta)$. Since the angular correlation function is practically a
smooth function, the numerical evaluations of derivatives of
$w(\theta)$ are stable enough once $w(\theta)$ is obtained.

Thus, in practice, only the self-correlation $w_{ii} = u(0)$ should be
evaluated in some way other than the equation (\ref{eq3-77}). This
should be done only once, so that the direct four-dimensional
numerical integration of equation (\ref{eq3-73}) is still reasonable.
Alternatively, since the Limber's equation is appropriate for the
self-correlation, one can use the equation (\ref{eq3-72}) with $\theta
= 0$,
\begin{equation}
  u(0) =
  \int_0^\infty dz H(z) n^2(z) D^2(z) b^2(z)
  \int_0^\infty \frac{k dk}{2\pi}
  W^2\left[k\theta_{\rm s} d_{\rm cA}(z)\right] P(k),
\label{eq3-78}
\end{equation}
which is two-dimensional integration and is faster to compute than the
direct four-dimensional integration.

Obtaining the one-dimensional function $u(\theta)$ by using
appropriate methods explained above, the correlation matrix $R_{ij}$
including the noise term is immediately constructed by (c.f.,
eq.[\ref{eq2-9}])
\begin{equation}
  R_{ij} = N_i N_j u(\theta_{ij}) + \sqrt{N_i N_j} K_{ij} + E_{ij},
\label{eq3-79}
\end{equation}
where $\theta_{ij}$ is the separation angle between cells $i$ and $j$,
$E_{ij}$ is the correlation matrix for other sources of noise
except the shot noise. The quantities $N_i$, $K_{ij}$ are defined by
\begin{eqnarray}
  N_i &=& \int \sin\theta d\theta d\phi
  K_i(\theta,\phi) n(\theta,\phi),
\label{eq3-80a}\\
  K_{ij} &=& \frac{1}{S_i S_j} \int \sin\theta d\theta d\phi
  K_i(\theta,\phi) K_j(\theta,\phi),
\label{eq3-80b}
\end{eqnarray}
where $n(\theta,\phi)$ is the mean number density of the 2D sample,
which can depend on the direction because of possibly inhomogeous
sampling, $K_i(\theta,\phi)$ is the smoothing kernel of the cell $i$
with an arbitrary normalization,
\begin{equation}
  S_i = 
  \int \sin\theta d\theta d\phi
  K_i(\theta,\phi).
\label{eq3-81}
\end{equation}
The normalized kernel $W_i(\theta,\phi)$ in equation (\ref{eq3-63}) is
given by $W_i(\theta,\phi) = K_i(\theta,\phi)/S_i$ (no sum over $i$).


\setcounter{equation}{0}
\section{
Application to the Likelihood Analysis
\label{sec4}
}

\subsection{
Cosmological Parameters of Simple Dependendence on the Correlation
Matrices
\label{sec4-1}
}

In the likelihood analysis, one needs to repeatedly calculate the
correlation matrices for models with various parameters. The methods
to calculate the correlation matrices described above are already
fast. However, when the dependence on some cosmological parameter is
linear or polynomial, the repeated compression of the correlation
matrix $R$ into a reduced matrix $C$ of equation (\ref{eq2-5}) is not
necessarily needed in changing that parameter. Generally, if the
dependence of a cosmological parameter $\theta$ on the correlation
matrix has the form,
\begin{equation}
  R = R^{(0)} + R^{(1)}\theta + R^{(2)}\theta^2 + \cdots,
\label{eq4-1}
\end{equation}
where $R^{(k)}$'s do not depend on $\theta$, then the reduced
correlation matrix of equation (\ref{eq2-5}) is given by
\begin{equation}
  C = C^{(0)} + C^{(1)}\theta + C^{(2)}\theta^2 + \cdots,
\label{eq4-2}
\end{equation}
where 
\begin{equation}
  C^{(k)} = P R^{(k)} P^{\rm T},
\label{eq4-3}
\end{equation}
and $P$ is the projection matrix of equation (\ref{eq2-4}). Therefore,
when the partial matrices $C^{(k)}$ are calculated fixing other
parameters, one can omit the projection of the huge matrix in changing
the particular parameter $\theta$. Instead, the reduced correlation
matrix $C$ is obtained simply by a summation of equation
(\ref{eq4-3}).

The obvious example of the cosmological parameter which has the
polynomial dependence on correlation matrices is the normalization
parameter of the mass power spectrum, ${\sigma_8}^2$, or $A_{\rm s}$.
The signal part of the correlation matrix, $\xi_{ij}$, or $w_{ij}$ is
always proportional to this parameter. The correlation matrix of a
redshift survey is written as
\begin{equation}
  R_{ij} =
  {\sigma_8}^2 N_i N_j \widehat{\xi}_{ij} +
  \sqrt{N_i N_j} K_{ij} + E_{ij}
\label{eq4-4}
\end{equation}
where $\widehat{\xi}_{ij}$ is the signal part of the correlation
matrix of equation (\ref{eq3-30}) with the normalization $\sigma_8 =
1$.

The dependence on the parameter $\beta$ at a fixed redshift is
polynomial. To see this, the expression of the correlation matrix
$\xi_{ij}$ of equations (\ref{eq3-30}) and
(\ref{eqa3-20a})--(\ref{eqa3-20i}) is re-arranged according to the
dependences on $\beta_i$ and $\beta_j$:
\begin{eqnarray}
  \xi_{ij} 
  &=&
  b_i b_j D_i D_j
  \left(
    \xi^{(0)}_{ij} + 
    \beta_i \xi^{(1)}_{ij} + \beta_j \xi^{(1)}_{ji} + 
    \beta_i \beta_j \xi^{(2)}_{ij}
  \right)
\label{eq4-5a}\\
  &=&
  D_i D_j
  \left(
    b_i b_j \xi^{(0)}_{ij} + 
    f_i b_j \xi^{(1)}_{ij} + f_j b_i \xi^{(1)}_{ji} + 
    f_i f_j \xi^{(2)}_{ij}
  \right)
\label{eq4-5b}
\end{eqnarray}
where $\xi^{(k)}_{ij}$ ($k=0,1,2$) do not depend on $\beta_i$,
$\beta_i$, or on $b_i$, $b_j$. The matrix $\xi^{(1)}_{ij}$ is not
symmetric in general. It is straightforward to obtain
$\xi^{(k)}_{ij}$. For convenience, below we give the explicit
representation in the case of $x_{ij} \ll |K|^{-1/2}$, which is
fulfilled in practice, since the curvature scale is observationally
more than five times larger than the Hubble scale as discussed just
above equation (\ref{eq3-32}). In this case we can put $|K|
{S_K}^2(x_{ij}) = 0$, $C_K(x_{ij}) = 1$ and $\widetilde{\theta}_{ij} =
\theta_{ij}$ in the equations (\ref{eqa3-20a})--(\ref{eqa3-20i}).
Therefore,
\begin{eqnarray}
&&
  \xi^{(0)}_{ij} = \xi^{(0)}_0(x_{ij}),
\label{eq4-6a}\\
&&
  \xi^{(1)}_{ij} = 
  \frac13 \xi^{(0)}_0(x_{ij}) + 
  A_{ij}\cos\gamma_{ij} \xi^{(1)}_1(x_{ij}) +
  \left(\cos^2\gamma_{ij} - \frac13 \right) \xi^{(1)}_2(x_{ij}),
\label{eq4-6b}\\
&&
  \xi^{(2)}_{ij} = 
  \frac{1}{15} \left(1 + 2\cos^2\theta_{ij}\right)
  \xi^{(0)}_0(x_{ij}) -
  \frac13 A_{ij}A_{ji} \cos\theta_{ij} \xi^{(1)}_0(x_{ij})
\nonumber\\
&&\qquad\quad +\,
  \frac15
  \left[
    A_{ij}
    \left(\cos\gamma_{ij} - 2\cos\gamma_{ji}\cos\theta_{ij}\right) +
    A_{ji}
    \left(\cos\gamma_{ji} - 2\cos\gamma_{ij}\cos\theta_{ij}\right)
  \right] \xi^{(1)}_1(x_{ij})
\nonumber\\
&&\qquad\quad -\,
  \frac17
  \left[
    \frac23 + \frac43\cos^2\theta_{ij} -
    \left(\cos^2\gamma_{ij} + \cos^2\gamma_{ji}\right) +
    4\cos\gamma_{ij}\cos\gamma_{ji}\cos\theta_{ij}
  \right] \xi^{(1)}_2(x_{ij})
\nonumber\\
&&\qquad\quad +\,
  A_{ij} A_{ji}
   \left(\cos\gamma_{ij}\cos\gamma_{ji} + \frac13 \cos\theta_{ij}\right)
   \xi^{(2)}_2(x_{ij})
\nonumber\\
&&\qquad\quad +\,
  \frac15
  \left[
    A_{ij}
    \left(
      5\cos\gamma_{ij}\cos^2\gamma_{ji} - \cos\gamma_{ij} +
      2\cos\gamma_{ji}\cos\theta_{ij}
    \right)
  \right.
\nonumber\\
&&\qquad\qquad\qquad
   \left. + \,
      A_{ji}
      \left(
         5\cos\gamma_{ji}\cos^2\gamma_{ij} - \cos\gamma_{ji} +
         2\cos\gamma_{ij}\cos\theta_{ij}
      \right)
   \right] \xi^{(2)}_3(x_{ij})
\nonumber\\
&&\qquad\quad +\,
   \frac17
   \left[
      \frac15 + \frac25\cos^2\theta_{ij} -
      \left(\cos^2\gamma_{ij} + \cos^2\gamma_{ji}\right) +
      4\cos\gamma_{ij}\cos\gamma_{ji}\cos\theta_{ij}
   \right.
\nonumber\\
&&\qquad\qquad\qquad  + \,
   \left.
      7\cos^2\gamma_{ij}\cos^2\gamma_{ji}
   \right] \xi^{(2)}_4(x_{ij}).
\label{eq4-6c}
\end{eqnarray}
The quantity $\xi^{(0)}_{ij}$ corresponds to the isotropic component
of the correlations, since it depends only on $x_{ij}$. The quantities
$\xi^{(1)}_{ij}$ and $\xi^{(2)}_{ij}$ are relevant to the distortions
by the peculiar velocity field.

When a fixed redshift $z_{\rm pivot}$ is arbitrarily chosen, the
correlation matrix is explicitly polynomial with respect to the
parameter $\beta_{\rm piv} \equiv \beta(z_{\rm pivot})$:
\begin{equation}
  \xi_{ij} =
  \left(b_{\rm piv} D_{\rm piv}\right)^2
  \widehat{b}_i \widehat{b}_j \widehat{D}_i \widehat{D}_j
  \left[
    \xi^{(0)}_{ij} + 
    \beta_{\rm piv} 
    \left(
      \widehat{\beta}_i \xi^{(1)}_{ij} + 
      \widehat{\beta}_j \xi^{(1)}_{ji}
    \right) + 
    {\beta_{\rm piv}}^2
    \widehat{\beta}_i \widehat{\beta}_j \xi^{(2)}_{ij}
  \right],
\label{eq4-7}
\end{equation}
where $b_{\rm piv} \equiv b(z_{\rm pivot})$ and $D_{\rm piv} \equiv
D(z_{\rm pivot})$ are the bias and the growth factor at a pivot
redshift. The relative evolutions with respect to $z_{\rm pivot}$ are
represented by quantities $\widehat{b}_i \equiv b(z_i)/b(z_{\rm
pivot})$, $\widehat{D}_i \equiv D(z_i)/D(z_{\rm pivot})$, and
$\widehat{\beta}_i \equiv \beta(z_i)/\beta(z_{\rm pivot})$. When the
redshift range of the survey is not significantly large, the relative
evolutions are small, and are not strongly varying functions of
redshift.

It is not hard to imagine that the typical bias $b_{\rm pivot}$ and
the overall normalization of the mass power spectrum at $z_{\rm
pivot}$, $\sigma_{\rm 8,piv} = D_{\rm piv} \sigma_8$, contribute to
the correlation matrix in similar way, and these parameters degenerate
in estimating the likelihood function, although the distortion terms
break this degeneracy to some extent. Since it is not advantageous to
deal with almost degenerated parameters separately, it might be
preferable to consider the normalization of the galaxy power spectrum
at $z_{\rm pivot}$, $\sigma_{\rm 8g,piv} \equiv b_{\rm piv}
\sigma_{\rm 8,piv} = b_{\rm piv} D_{\rm piv} \sigma_8$, as an
independent parameter instead of the unobservable $\sigma_8$. We
introduce the normalized correlations, $\widehat{\xi}^{(k)}_{ij}$
($k=0,1,2$), which are calculated from the power spectrum with a
normalization $\sigma_8 = 1$, so that $\xi^{(k)}_{ij} = {\sigma_8}^2
\widehat{\xi}^{(k)}_{ij}$. The correlation matrix of equation
(\ref{eq4-7}) is then reduced to
\begin{equation}
  R_{ij} =
  {\sigma_{\rm 8g,piv}}^2
  N_i N_j
  \widehat{b}_i \widehat{b}_j \widehat{D}_i \widehat{D}_j
  \left[
    \widehat{\xi}^{(0)}_{ij} + 
    \beta_{\rm piv} 
    \left(
      \widehat{\beta}_i \widehat{\xi}^{(1)}_{ij} + 
      \widehat{\beta}_j \widehat{\xi}^{(1)}_{ji}
    \right) + 
    {\beta_{\rm piv}}^2
    \widehat{\beta}_i \widehat{\beta}_j \widehat{\xi}^{(2)}_{ij}
  \right]
  + \sqrt{N_i N_j} K_{ij} + E_{ij},
\label{eq4-8}
\end{equation}
(no sum over $i$, $j$). Thus the normalization of the galaxy power
spectrum at the pivot redshift, $\sigma_{\rm 8g,piv}$, and the
redshift-distortion parameter at the pivot redshift, $\beta_{\rm
piv}$, are independent parameters both of which polynomially
contribute to the correlation matrix. The parameters $\sigma_8$ and
$b_{\rm piv}$ are dependent parameters through the relations
\begin{eqnarray}
  \sigma_8
  &=&
  \frac{\sigma_{\rm 8g,piv}\beta_{\rm piv}}
    {f(z_{\rm pivot}) D(z_{\rm pivot})},
\label{eq4-9-1a}\\
  b_{\rm piv} &=& \frac{f(z_{\rm pivot})}{\beta_{\rm piv}}.
\label{eq4-9-1b}
\end{eqnarray}
In similar ways, one can arbitrary choose two independent parameters
out of $\sigma_8$, $b_{\rm piv}$, $\beta_{\rm piv}$, and $\sigma_{\rm
8g,piv}$, depending on individual analysis.

Each term in the equation (\ref{eq4-8}) is separately projected to the
reduced matrices as explained above. As detailed in
Appendix~\ref{app1}, the projection matrix $P$ whitens the noise
correlation matrix, and the reduced correlation matrix $C = P R P^{\rm
T}$ is given by
\begin{equation}
  C_{nm} =
  {\sigma_{\rm 8g,piv}}^2
  \left(
    C^{(0)}_{nm} +  \beta_{\rm piv} C^{(1)}_{nm} +
    {\beta_{\rm piv}}^2 C^{(2)}_{nm}
  \right)
  + \delta_{nm},
\label{eq4-9}
\end{equation}
where
\begin{eqnarray}
&&
  C^{(0)}_{nm} =
  \sum_{i,j}
  P_{ni} P_{mj} N_i N_j
  \widehat{b}_i \widehat{b}_j \widehat{D}_i \widehat{D}_j
  \widehat{\xi}^{(0)}_{ij},
\label{eq4-10a}\\
&&
  C^{(1)}_{nm} =
  \sum_{i,j}
  P_{ni} P_{mj} N_i N_j
  \widehat{D}_i \widehat{D}_j
  \left(
    \widehat{f}_i \widehat{b}_j \widehat{\xi}^{(1)}_{ij} +
    \widehat{f}_j \widehat{b}_i \widehat{\xi}^{(1)}_{ij}
  \right),
\label{eq4-10b}\\
&&
  C^{(2)}_{nm} =
  \sum_{i,j}
  P_{ni} P_{mj} N_i N_j
  \widehat{D}_i \widehat{D}_j
  \widehat{f}_i \widehat{f}_j \widehat{\xi}^{(2)}_{ij}.
\label{eq4-10c}
\end{eqnarray}
Once the projections of the matrices in the equations
(\ref{eq4-10a})--(\ref{eq4-10c}) are obtained for a given set of
parameters except $\sigma_{\rm 8g,piv}$ and $\beta_{\rm piv}$, the
reduced correlation matrix $C_{nm}$ is immediately derived by the
equation (\ref{eq4-9}) for any values of $\sigma_{\rm 8g,piv}$ and
$\beta_{\rm piv}$ without projecting the huge matrix again. Other
cosmological parameters, including $\mOmega_{\rm M0}$, $\mOmega_{\rm
Q0}$, and parameters which parametrize the evolution of the bias,
$\widehat{b}(z)$, and of the equation of state for the dark energy,
$w(z)$, are non-polynomially dependent on the correlation matrix
$R_{ij}$. The projection operation should be performed for each set of
these non-polynomial parameters.

In shallow redshift surveys, however, the evolutionary effects of the
cosmological quantities above are small. In this case, the natural
choice of the pivot redshift is $z_{\rm pivot} = 0$ so that
$\sigma_{\rm 8g,piv} = \sigma_{\rm 8g}$ and $\beta_{\rm piv} =
\beta_0$. One can also approximately set $\widehat{D} = 1$,
$\widehat{f} = 1$, $\widehat{b} = 1$, and $H(z) = H_0$ in the formula
of the correlation matrix ($H_0$ does not explicitly affect the
correlation matrix in redshift space for a given power spectrum, only
through the power spectrum). The only non-polynomial parameters are
those which parametrize the power spectrum, since, with the good
approximation, the distances are proportional to the redshifts and the
geometry is flat. Therefore, in this case, the elements
$\widehat{\xi}^{(k)}_{ij}$ are fully determined where any dependence
on cosmological parameters is only through the power spectrum. The
representation of equations (\ref{eq4-10a})--(\ref{eq4-10c}) are
simplified as
\begin{eqnarray}
&&
  C^{(0)}_{nm} =
  \sum_{i,j}
  P_{ni} P_{mj} N_i N_j \widehat{\xi}^{(0)}_{ij},
\label{eq4-11a}\\
&&
  C^{(1)}_{nm} =
  \sum_{i,j}
  P_{ni} P_{mj} N_i N_j
  \left(
    \widehat{\xi}^{(1)}_{ij} +
    \widehat{\xi}^{(1)}_{ji}
  \right),
\label{eq4-11b}\\
&&
  C^{(2)}_{nm} =
  \sum_{i,j}
  P_{ni} P_{mj} N_i N_j
  \widehat{\xi}^{(2)}_{ij},
\label{eq4-11c}
\end{eqnarray}
which are only dependent on the shape of the power spectrum.

\subsection{
Band Power Estimation of the Power Spectrum
\label{sec4-2}
}

The correlation matrix is always linearly dependent on the power
spectrum as long as the linear regime is considered. Therefore,
decomposing the power spectrum into band powers can be used to
straightforwardly estimate the power spectrum, without parameterizing
the shape of the power spectrum \citep{teg02}. Below we briefly
explain the band power estimation within the context of our method.

The power spectrum is decomposed into band powers as
\begin{equation}
  P(k) = \sum_{p=1}^{n_p} A_p \widehat{P}_p(k),
\label{eq4-20}
\end{equation}
where $\widehat{P}_p(k)$ is the band-power base function which has
support near the wavenumber $k_p$, and $A_p$ is the power of the band
$p$. The choice of the band-power base function is not unique, as long
as the power spectrum is parameterized by linear parameters $A_p$. The
simplest choice is the piecewise constant functions,
\begin{equation}
  \widehat{P}_p(k) = \mTheta(k_{p+1} - k)\mTheta(k - k_p).
\label{eq4-21}
\end{equation}
Another choice is the piecewise linear functions,
\begin{equation}
  \widehat{P}_p(k) =
  \frac{k-k_{p-1}}{k_p - k_{p-1}}
  \mTheta(k_p - k)\mTheta(k -  k_{p-1}) +
  \frac{k_{p+1} - k}{k_{p+1} - k_p}
  \mTheta(k_{p+1} - k)\mTheta(k -  k_p).
\label{eq4-22}
\end{equation}
One can construct more complex bases by optimizing the band powers
\citep[e.g., see][]{ham97,ham00}. Each band contributes linearly to the
reduced correlation matrix $C_{nm}$. We denote the contribution of
each band $p$ to the matrices $\xi^{(a)}_{ij}$ $(a=0,1,2)$ of
equations (\ref{eq4-6a})--(\ref{eq4-6c}) by $\xi^{(a,p)}_{ij}$ such
that
\begin{equation}
  \xi^{(a)}_{ij} = 
  \sum_{p=1}^{n_p} A_p \xi^{(a,p)}_{ij}.
\label{eq4-23}
\end{equation}
Then, analogously to the equations (\ref{eq4-9})--(\ref{eq4-10c}), the
reduced correlation matrix is given by
\begin{equation}
  C_{nm} =
  \sum_{a=0}^{2}
  \sum_{p=1}^{n_p} A_p
  {\beta_{\rm piv}}^a C^{(a,p)}_{nm}
  + \delta_{nm},
\label{eq4-24}
\end{equation}
where
\begin{eqnarray}
&&
  C^{(0,p)}_{nm} =
  \sum_{i,j}
  P_{ni} P_{mj} N_i N_j
  \widehat{b}_i \widehat{b}_j \widehat{D}_i \widehat{D}_j
  \xi^{(0,p)}_{ij},
\label{eq4-25a}\\
&&
  C^{(1,p)}_{nm} =
  \sum_{i,j}
  P_{ni} P_{mj} N_i N_j \widehat{D}_i \widehat{D}_j
  \left(
    \widehat{f}_i \widehat{b}_j \xi^{(1,p)}_{ij} +
    \widehat{f}_j \widehat{b}_i \xi^{(1,p)}_{ji}
  \right),
\label{eq4-25b}\\
&&
  C^{(2,p)}_{nm} =
  \sum_{i,j}
  P_{ni} P_{mj} N_i N_j \widehat{D}_i \widehat{D}_j
  \widehat{f}_i \widehat{f}_j \xi^{(2,p)}_{ij}.
\label{eq4-25c}
\end{eqnarray}

The matrices $C^{(a,p)}_{nm}$ still depend on the cosmological
parameters because of the evolutionary effects. However, in case of
shallow surveys, they become independent on any parameters at all.
Once the matrices $C^{(a,p)}_{nm}$ are obtained, the dependencies of
the correlation matrix on parameters $a_p$ and $\beta_0$ are
explicitly polynomial.


\subsection{
Quadratic Estimator
\label{sec4-3}
}

When the number of parameters to be estimated is large, it is not
realistic to calculate the likelihood function on every mesh point in
the parameter space. When the behavior of the likelihood function is
completely unknown, there are no safe methods to maximize the
likelihood in parameter space. In the present application in
cosmology, the correlation matrix smoothly varies with cosmological
parameters in general. In this case, the likelihood function near a
peak is approximated as a Gaussian. Therefore, as long as we
approximately know the true cosmological parameters, one can reach the
location of maximun likelihood by a few iterative steps using the
Newton-Raphson method. This approach is taken by \citet{bon98}
(hereafter, BJK) in estimating the power spectrum of the cosmic
microwave background (CMB). The same method is also applied to the
analysis of the 2-dimensional galaxy survey \citep{hut01,efs01}. The
purpose of this section is to give an aspect of applying BJK's method
to the present analysis.

Following BJK, the likelihood function of equation (\ref{eq2-0}),
${\cal L}(\{\theta_\alpha\}) \equiv P(D_{\rm reduced}|\mTheta)$ is
approximated by a Gaussian:
\begin{equation}
  \ln {\cal L}\left(\{\theta_\alpha + \delta\theta_\alpha\}\right) =
  \ln {\cal L}\left(\{\theta_\alpha\}\right) +
  \sum_\alpha
  \frac{\partial \ln {\cal L}}{\partial \theta_\alpha}
  \delta \theta_\alpha +
  \frac12 \sum_{\alpha,\beta}
  \frac{\partial\ln{\cal L}}
    {\partial \theta_\alpha \partial \theta_\beta}
  \delta \theta_\alpha \delta \theta_{\beta}
\label{eq4-40}
\end{equation}
In this approximation, the parameters which maximize the likelihood is
directly solved as
\begin{equation}
  \delta \theta_\alpha =
  - \sum_{\beta} 
  \left(
    \frac{\partial^2 \ln {\cal L}}
      {\partial\theta_\alpha \partial\theta_\beta}
  \right)^{-1}
  \frac{\partial\ln{\cal L}}{\partial\theta_\beta},
\label{eq4-41}
\end{equation}
where first- and second-derivatives of the log likelihood function is
explicitly calculated from equation (\ref{eq2-0}), resulting in
\begin{eqnarray}
&&
  \frac{\partial\ln{\cal L}}{\partial\theta_\alpha} =
  \frac12 {\rm Tr}
  \left[
    \left(\bfm{B} \bfm{B}^{\rm T} - C\right)
    \left(C^{-1} C_{,\alpha} C^{-1}\right)
  \right]
\label{eq4-42a}\\
&&
  \frac{\partial^2\ln{\cal L}}
    {\partial\theta_\alpha\partial\theta_\beta} =
  \frac12
  {\rm Tr}
  \left[
    \left(\bfm{B} \bfm{B}^{\rm T} - C\right)
    \left(
      C^{-1} C_{,\alpha\beta} C^{-1} -
      C^{-1} C_{,\alpha} C^{-1} C_{,\beta} C^{-1} -
      C^{-1} C_{,\beta} C^{-1} C_{,\alpha} C^{-1}
    \right)
  \right.
\nonumber\\
&&\qquad\qquad\qquad\quad
  \left.
    - C^{-1} C_{,\alpha} C^{-1} C_{,\beta}
  \right],
\label{eq4-42b}
\end{eqnarray}
where $C$ is the correlation matrix and standard notations,
$C_{,\alpha} = \partial C/\partial\theta_\alpha$, etc., are employed.
Instead of intensive calculation of the curvature matrix of equation
(\ref{eq4-42b}), the second-order derivatives are replaced by an
expectation value, which is equivalent to the Fisher matrix:
\begin{equation}
  F_{\alpha\beta} \equiv -
  \left\langle
    \frac{\partial^2\ln{\cal L}}
    {\partial\theta_\alpha\partial\theta_\beta}
  \right\rangle =
  \frac12 {\rm Tr}\left[C^{-1} C_{,\alpha} C^{-1} C_{,\beta}\right].
\label{eq4-43}
\end{equation}
In this way, we obtain the BJK's quadratic estimator,
\begin{equation}
  \delta \theta_\alpha =
  \frac12 \sum_{\beta} 
  \left(F^{-1}\right)_{\alpha\beta}
  {\rm Tr}
  \left[
    \left(\bfm{B} \bfm{B}^{\rm T} - C\right)
    \left(C^{-1} C_{,\beta} C^{-1}\right)
  \right],
\label{eq4-44}
\end{equation}
which proves to be very useful. Only few iterations are needed to
reach the local maxima of the likelihood function when the band power
of the CMB spectrum is estimated by this method (BJK). Since the
projection matrix $P$ of the projected correlation matrix $C = P R
P^{\rm T}$ of equation (\ref{eq2-5}) is fixed throughout the
likelihood maximization, the derivatives $C_{,\alpha}$ are
straightforwardly calculated by numerically differentiating the
correlation matrix $R$: $C_{,\alpha} = P R_{,\alpha} P^{\rm T}$. For
the cosmological parameters with polynomial influence on the
correlation matrix, as considered in the section
\ref{sec4-1}, the derivatives $C_{,\alpha}$ are obviously given
without any numerical differentiation.




\setcounter{equation}{0}
\section{
Conclusions and Discussion
\label{sec6}
}

Technical details of the methods in constructing the correlation
matrix in redshift space, which are essential in direct likelihood
analysis of the large-scale structure, have been presented. Making use
of the most general analytic formula of the linear two-point
correlation function in redshift space given by \citet{mat00}, a fast
procedure to construct correlation matrices in redshift space has been
explicitly provided. This procedure can be used for deep redshift
surveys, as well as shallow surveys. The finger-of-God effects are not
significant as long as the nonlinear modes are excluded by truncating
the KL-modes. We have also given fast methods to produce correlation
matrices in projected samples on the sky in which spatial and
spherical curvature effects are taken into account. The Limber's
equation is not appropriate for large separations. Some parameters
polynomially depend on the correlation matrix. Those parameters are
evidently simple to analyze. When the number of parameters are large,
the quadratic estimator is a promising way to maximize the likelihood
function.

The purpose of this paper is to describe technical details of
computationally quick construction of correlation matrices. Parts of
the present methods have been successfully applied to the real data,
including the Las Campanas Redshift Survey and SDSS
\citep{mat00b,sza03,pop04}. So far the application has been restricted
to the shallow samples and the projected samples of galaxies. One of
the final targets of the present methods is the application to the
large-scale redshift surveys, which are deep and wide enough in
redshift space. In this respect, our methods are effective in the
analysis of the ongoing SDSS luminous red galaxy (LRG) sample
\citep{eis01}, in which $\sim 10^5$ galaxies are catalogued with
redshift range of $z\sim$ 0.2--0.5 over $\sim 10^4$ square degrees on
the sky. Since this sample is dominated by the shot-noise on smaller
scales, the KL transform is essential to maximally extract the
cosmological information. The redshifts in the LRG sample are not very
shallow as in the main galaxy sample, in which $z \simlt 0.2$, the
evolutionary effects on the clustering is detectable, and therefore
geometry of the universe, nature of dark energy, etc., can be finely
constrained because of the cosmological redshift-space distortion
effects \citep{mat01,mat02,mat03,bla03,seo03}. In next-generation
redshift surveys beyond the SDSS, the present methods will provide a
unique technique in cosmological analyses of the survey data. In large
and deep redshift surveys of the future, the linear regime will be
more focused on, and the shot noise will be more severe than the
present-day surveys. The advantage of the methods developed in this
paper is much greater in those future surveys than the past shallow
surveys.


\acknowledgements

We thank Daniel Eisenstein and Tamas Budavari for discussion. TM
acknowledges support from the Ministry of Education, Culture, Sports,
Science, and Technology, Grant-in-Aid for Encouragement of Young
Scientists, 15740151, 2003. AS acknowledges support from grants NSF
AST-9802 980 and NASA LTSA NAG-53503.


\appendix

\section{Construction of the Projection Matrix by the KL Eigenmodes
\label{app1}}

In this appendix, the construction of the projection matrix $P$ of
equation (\ref{eq2-4}) which reduces the dimensionality of the data
space is reviewed, based on the method of \citet{vog96}.

To capture maximum signals with minimum noise, one needs to keep the
modes with the high S/N ratios and discard the modes with the low S/N
ratios. For this purpose, we decompose the correlation matrix $R$ into
the signal part $S$ plus the noise part $N$, $R = S + N$, assuming
signals and noises are mutually uncorrelated. To normalize the noise
correlations, we consider a linear transformation,
\begin{equation}
   \bfm{d}' = Q \bfm{d},
\label{eqa10}
\end{equation}
where $Q$ is a non-degenerate $N\times N$ matrix, $\det Q \neq 0$,
which is not neccesarily a symmetric matrix. The transformation $Q$ is
chosen so that the noise correlation matrix in the new data vector
$\bfm{d}'$ is transformed to an identity matrix:
\begin{equation}
   N' = Q N Q^{\rm T} = I,
\label{eqa11}
\end{equation}
This transformation is referred to by {\em prewhitening} \citep{vog96}.

However, the choice of the prewhitening matrix $Q$ is not unique when
the noise correlation matrix is not diagonal. For example, one can
first diagonalize the noise correlation matrix $N$ and then multiply a
weighted diagonal matrix:
\begin{equation}
   Q = Q_2 Q_1,
\label{eqa12}
\end{equation}
where $Q_1$ is an orthogonal matrix which diagonalize the noise matrix
$N$:
\begin{equation}
   Q_1 N {Q_1}^{\rm T} =
   {\rm diag.}({\sigma_1}^2, {\sigma_2}^2, \ldots {\sigma_N}^2),
\label{eqa13}
\end{equation}
and $Q_2$ is a diagonal matrix weighted by the inverse of the square
root of the noise
eigenvalues:
\begin{equation}
   Q_2 = 
  {\rm diag.}({\sigma_1}^{-1}, {\sigma_2}^{-1}, \ldots {\sigma_N}^{-1}).
\label{eqa14}
\end{equation}
This choice is not the only possibility. Another example of the
transformation $Q$ is given by the Cholesky decomposition of the noise
matrix, $N = L L^{\rm T}$, where $L$ is a lower triangular matrix. The
choice $Q = L^{-1}$ also satisfies the equation (\ref{eqa11}). Unless
the noise matrix has diagonal form in first place, the matrix $Q_2
Q_1$ is not a triangular matrix and the matrix $L^{-1}$ is a lower
triangular matrix so that $Q_2 Q_1 \neq L^{-1}$ in general.

Once the noise matrix is normalized to an identity matrix by a certain
choice of $Q$, the correlation matrix of equation (\ref{eq2-3}) is
transformed to
\begin{equation}
   R' = Q S Q^{\rm T} + I.
\label{eqa15}
\end{equation}
Now, to obtain the {\em statistically orthogonal} set of data vectors,
the prewhitened data is rotated so that the correlation matrix $R'$ is
diagonalized, solving the eigenvalue equation,
\begin{equation}
   R'\bfm{\mPsi}_n = \lambda_n'\bfm{\mPsi}_n.
\label{eqa16}
\end{equation}
Since the diagonalization of the matrix $R'$ keeps the noise term as
the identity matrix, the noise contribution to an eigenvalue
$\lambda'$ is always unity. Therefore, the larger the eigenvalue is,
the larger the signal-to-noise ratio of the corresponding eigenmode
is. Sorting the eigenvalues in decreasing order,
\begin{equation}
   \lambda_1' \geq \lambda_2' \geq \cdots \geq \lambda_M'
   \geq \cdots \geq \lambda_N',
\label{eqa17}
\end{equation}
and discarding the modes with low eigenvalues of
$\lambda_{M+1}',\ldots,\lambda_{N}'$, the natural choice of the
projected data is
\begin{equation}
   B_n = \bfm{\mPsi}_n^{\rm T} \bfm{d}' = 
   \bfm{\mPsi}_n^{\rm T} Q \bfm{d}
\label{eqa18}
\end{equation}
where $n \leq M$, and the eigenvectors satisfy the orthonormality,
\begin{equation}
   \bfm{\mPsi}_n^{\rm T}\bfm{\mPsi}_m = \delta_{nm}.
\label{eqa19}
\end{equation}
Employing the projection matrix of the whitened data,
\begin{equation}
   P' = 
   \left(
     \bfm{\mPsi}_1, \bfm{\mPsi}_2, \ldots, \bfm{\mPsi}_M
   \right)^{\rm T},
\label{eqa20}
\end{equation}
the optimal projection matrix $P$ is given by
\begin{equation}
   P = P' Q
\label{eqa21}
\end{equation}

The eigenvalue $\lambda_n'$ corresponds to ``the signal-to-noise ratio
plus unity'' for each mode, since one can notice that the matrix $Q$ is
the ``square root'' of the noise matrix (c.f., eq.[\ref{eqa11}]).
Therefore, the quantities $\lambda_n \equiv \lambda_n' - 1$ correspond
to the ``signal-to-noise'' eigenvalues. This is more clearly shown
by the fact that the eigenvalue equation (\ref{eqa12}) is equivalent
to the generalized eigenvalue equation
\begin{equation}
   S \bfm{\mPhi}_n = \lambda_n N \bfm{\mPhi}_n,
\label{eqa22}
\end{equation}
where $\bfm{\mPhi}_n = Q^{\rm T}\bfm{\mPsi}_n$ is the signal-to-noise
eigenmode, which is introduced by \citet{bon95} in the analysis of the
CMB data. The projection matrix $P$ is simply given by
\begin{equation}
   P =
   \left(
     \bfm{\mPhi}_1, \bfm{\mPhi}_2, \ldots, \bfm{\mPhi}_M
   \right)^{\rm T}.
\label{eqa23}
\end{equation}
In this representation, it is explicitely seen that the choice of the
projection matrix $P$ is independent on the choice of the prewhitening
matrix $Q$. The signal-to-noise eigenvectors $\bfm{\mPhi}$ are no longer
orthogonal, but are normalized by
\begin{equation}
   \bfm{\mPhi}_n^{\rm T} N \bfm{\mPhi}_m = \delta_{nm}.
\label{eqa24}
\end{equation}

\begin{figure}[ht]
\epsscale{0.8} \plotone{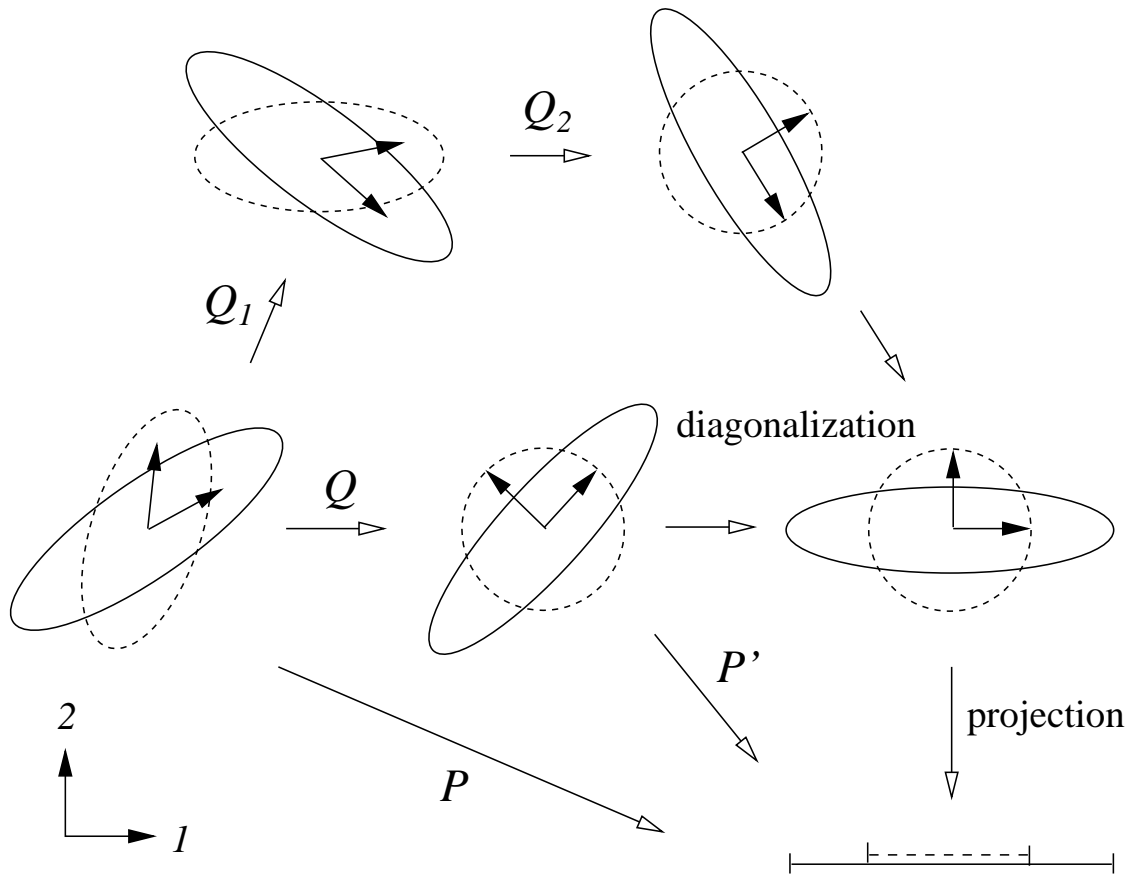} \figcaption[f_proj.eps]{ A
schematic explanation of the projection by the S/N eigenmodes, or KL
eigenmodes. Solid ellipses represent the signal correlations and
dashed ellipses represent the noise correlations in two-dimensional
data space. The directions of the S/N eigenmodes are indicated by
black arrows. The matrix $Q_1$ diagonalizes the noise correlations,
and $Q_2$ rescales the data space to obtain the prewhitened noise
matrix. Although there is not a unique way to choose a prewhitening
matrix $Q$, the resulting projection matrix $P$ is independent of the
choice. The generalized eigenvalue equation directly gives the
projection matrix $P$. The original data space is projected onto the
reduced data space, retaining only modes with the highest S/N ratio.
\label{fig-proj}}
\end{figure}

A schematic explanation of obtaining the S/N eigenmodes is given in
Figure~\ref{fig-proj}.
The solid ellipses represent the signal correlations and the dashed
ellipses represent the noise correlations in two-dimensional data
space. The S/N eigenmodes, or the KL modes, the directions of which is
indicated by black arrows, determines the direction of the projection
to reduce the dimension of the data space, retaining the maximal
signals-to-noise ratio.

In the above procedure, the correlation matrix of the reduced data
$\bfm{B}$ is a diagonal matrix since it is obtained by diagonalization
of the correlation matrix of the prewhitened data $\bfm{d}'$.
Therefore, the reduced data is statistically orthogonal:
\begin{equation}
   C_{nm} = \lambda_n' \delta_{nm}
\label{eqa25}
\end{equation}
where $\lambda_n' = \bfm{\mPsi}_n^{\rm T} R' \bfm{\mPsi}_n =
\bfm{\mPhi}_n^{\rm T} R \bfm{\mPhi}_n = \lambda_n + 1 = \langle
(B_n)^2 \rangle - {\langle B_n \rangle}^2$. The above orthogonality is
achieved only when the true cosmological model of the correlation
matrix $R$ is known. In practice, the true parameters of the
cosmological model are the object of study. Therefore the
orthogonality of $C_{nm}$ is only approximate and is never assumed in
our likelihood analysis.

\section{The Linear Formula of the Correlation Function in Redshift
  Space with the Curvature Effects
\label{app2}}

In this appendix, we summarize the most general expression for the
linear correlation function derived by \citet{mat00}. The
high-redshift effects, the wide-angle effects, the peculiear velocity
effects, and the selection effects are all included.

To describe the formula, we need the orthonormal modes of the
Laplacian in the spatial section of the RW metric. That is, the
orthonormal modes of the equation
\begin{equation}
  (\triangle + k^2 - K) Z = 0,
\label{eqa2-1}
\end{equation}
prove to be useful. In the spatial section of the RW metric of
equation (\ref{eq3-1}), the above equation reduces to 
\begin{equation}
  \frac{1}{{S_K}^2(x)}
  \left[
    \frac{\partial}{\partial x}
    \left({S_K}^2(x)\frac{\partial Z}{\partial x}\right) +
    \frac{1}{\sin\theta} \frac{\partial}{\partial\theta}
    \left(\sin\theta\frac{\partial Z}{\partial\theta}\right) +
    \frac{1}{\sin^2\theta}\frac{\partial^2 Z}{\partial\phi^2}
  \right]
  + (k^2 - K) Z = 0.
\label{eqa2-2}
\end{equation}
Separating the variable $Z$ into a radial part and an angular part,
the orthonormal modes have the form $X_l(k,x) Y_l^m(\theta,\phi)$,
where $Y_l^m$ are the spherical harmonics. The orthonormality and the
completeness of the spherical harmonics are useful:
\begin{eqnarray}
&&
  \int \sin\theta d\theta d\phi
  {Y_l^m}^*(\theta,\phi) Y_{l'}^{m'}(\theta,\phi) =
  \delta_{ll'} \delta_m^{m'},
\label{eqa2-2-1a}\\
&&
  \sum_{l=0}^\infty \sum_{m=-l}^l
  {Y_l^m}^*(\theta,\phi) Y_l^m(\theta',\phi') =
  \frac{\delta(\theta - \theta') \delta(\phi - \phi')}{\sin\theta}.
\label{eqa2-2-1b}
\end{eqnarray}
The radial part $X_l$ satisfies the differential equation
\begin{equation}
   \frac{1}{{S_K}^2(x)} \frac{\partial}{\partial x}
   \left(
      {S_K}^2(x) \frac{\partial X_l}{\partial x}
   \right) +
   \left[
      k^2 - \frac{l(l+1)}{{S_K}^2(x)}
   \right] X_l = 0.
\label{eqa2-3}
\end{equation}
Putting $f={S_K}^{1/2} X_l$ and $dz = S_K dx$, this equation reduces
to the associated Legendre differential equation. The solutions of
this equation are given by the conical function, the Bessel function,
and the toroidal function, for negative, zero, and positive
curvatures, respectively \citep{har67,wil83,mat00}. These solutions
which are regular at the origin $x=0$ are explicitly represented by
\begin{equation}
  X_l(k,x) =
  \frac{{S_K}^l(x)}{\sqrt{N_l(k)}}
  \left(
    \frac{1}{S_K(x)}\frac{\partial}{\partial x}
  \right)^l
  \left(\frac{\sin k x}{S_K(x)}\right), 
\label{eqa2-4}
\end{equation}
where 
\begin{equation}
  N_l(k) = 
  \prod_{j=0}^l (k^2 - j^2 K) = 
  k^2 (k^2 - K) (k^2 - 4 K)
    \cdots (k^2 - l^2 K),
\label{eqa2-5}
\end{equation}
are the normalization constants. For $x \rightarrow 0$, $X_l$ behaves
like $x^l$ with appropriate constants, and thus $X_l(k,0) =
\delta_{l0}$ \citep{har67}. For the flat universe, $K=0$, the equation
(\ref{eqa2-4}) is simply given by the spherical Bessel function,
$X_l(k,x) = (-1)^l j_l(kx)$. For $K \leq 0$, $k$ takes all positive
values. In the case of positive curvature, $K > 0$, the 3-space is
periodic and only discrete values $k = (l+1)\sqrt{K}, (l+2)\sqrt{K},
(l+3)\sqrt{K}, \ldots$ are allowed. In the latter case, the functions
(\ref{eqa2-4}) are reduced to the Gegenbauer polynomials. The
following recursion relations hold \citep{mat00}:
\begin{equation}
  \sqrt{k^2 - l^2 K} X_{l-1}(k,x) + 
  (2l + 1) \frac{C_K(x)}{S_K(x)} X_l(k,x) +
  \sqrt{k^2 - (l+1)^2 K} X_{l+1}(k,x) = 0
\label{eqa2-5-1}
\end{equation}

The orthonormal and completeness relations are
\begin{eqnarray}
&&
   4\pi \int_0^\infty dx\,
      {S_K}^2(x) X_l(k,x)  X_l(k',x) 
   = \frac{2\pi^2 \delta(k - k')}{k^2},
\label{eqa2-6a}\\
&&
   \int_0^\infty \frac{k^2 dk}{2\pi^2} X_l(k,x)  X_l(k,x')
   = \frac{ \delta(x - x')}{4\pi {S_K}^2(x)},
\label{eqa2-6b}
\end{eqnarray}
for $K \leq 0$, and 
\begin{eqnarray}
&&
   4\pi \int_0^\infty dx\,
      {S_K}^2(x) X_l(k_n,x)  X_l(k_{n'},x) 
   =
   \frac{2 \pi^2 \delta_{nn'}}{{k_n}^2},
\label{eqa2-6c}\\
&&
   \frac{\sqrt{K}}{2\pi^2} \sum_{n=l+1}^\infty
   {k_n}^2 X_l(k_n,x)  X_l(k_n,x')
   =
   \frac{\delta(x - x')}{4\pi {S_K}^2(x)},
\label{eqa2-6d}
\end{eqnarray}
for $K > 0$, where $k_n \equiv n\sqrt{K}$. In the following, only the
case of $K\leq 0$ is explicitly presented. The case of $K > 0$ is
easily obtained by discretizing the variable $k \rightarrow k_n$ and
by an interpretation of the integrals $\int_0^\infty dk \rightarrow
\sqrt{K} \sum_{n=l+1}^\infty$, and by $\delta(k - k') \rightarrow
\delta_{nn'}$ etc. For our purpose, we need the explicit forms of
$X_l$ only up to $l = 4$:
\begin{eqnarray}
  X_0 &=& \frac{\sin kx}{kS_K(x)},
\label{eq3-12a}\\
   X_1 &=& 
   \frac{1}{\sqrt{N_1(k)}\; {S_K}^2(x)}
   \left[ - {C_K}(x)\sin kx
       + k {S_K}(x)\cos kx\right],
\label{eq3-12b}\\
   X_2 &=& 
   \frac{1}{\sqrt{N_2(k)}\; {S_K}^3(x)}
   \left\{
      \left[
         3 - (k^2 + 2K){S_K}^2(x)
      \right]
      \sin kx - 3k {S_K}(x){C_K}(x)\cos kx
   \right\},
\label{eq3-12c}\\
   X_3 &=& 
   \frac{1}{\sqrt{N_3(k)}\; {S_K}^4(x)}
   \left\{
      {C_K}(x)
      \left[
         - 15 + 6 (k^2 + K) {S_K}^2(x)
      \right]
      \sin kx
   \right.
\nonumber\\
&& \qquad\qquad\qquad\qquad\qquad +\,
   \left.
      k{S_K}(x)
      \left[
         15 - (k^2 + 11K) {S_K}^2(x)
      \right]
      \cos kx
   \right\},
\label{eq3-12d}\\
   X_4 &=&
   \frac{1}{\sqrt{N_4(k)}{S_K}^5(x)}
   \left\{
      \left[
         105 - 15(3k^2 + 8K) {S_K}^2(x) +
         (k^4 + 35K k^2 + 24K^2) {S_K}^4(x)
      \right]
      \sin kx
   \right.
\nonumber\\
&& \qquad\qquad\qquad\qquad -\,
   \left.
      k {S_K}(x)
      \left[
         105 - 10(k^2 + 5K) {S_K}^2(x)
      \right]
      \cos k x
   \right\},
\label{eqa3-12e}
\end{eqnarray}
where the function
\begin{eqnarray}
   C_K(x) \equiv
   \frac{d S_K(x)}{dx} =
   \left\{
   \begin{array}{ll}
      \displaystyle
         \cosh\left(x\sqrt{-K}\right), & K < 0,
      \vspace{0.1cm} \\
      \displaystyle
      1, & K = 0,
      \vspace{0.1cm} \\
      \displaystyle
         \cos\left(x\sqrt{K}\right), & K > 0,
   \end{array}
   \right.
\label{eq3-13}
\end{eqnarray}
has the properties
\begin{eqnarray}
&&
   {C_K}^2(x) + K {S_K}^2(x) = 1,
\label{eq3-14a}\\
&&
   \frac{dC_K(x)}{dx} = - K S_K(x).
\label{eq3-14b}
\end{eqnarray}

Expanding the density contrast $\delta(x, \theta, \phi)$ in terms of
the normal modes,
\begin{eqnarray}
&&
  \delta(x,\theta,\phi) = 
  \sum_{l=0}^\infty \sum_{m=-l}^l
  \int \frac{k^2 dk}{2\pi^2} \widetilde{\delta}_{lm}(k)
  X_l(k,x) Y_l^m(\theta,\phi),
\label{eqa3-15a}\\
&&
  \widetilde{\delta}_{lm}(k) = 
  4\pi \int dx {S_K}^2(x) \int \sin\theta d\theta d\phi 
  \delta(x,\theta,\phi) 
  X_l(k,x) {Y_l^m}^*(\theta,\phi),
\label{eqa3-15b}
\end{eqnarray}
the power spectrum $P(k)$ is defined by 
\begin{equation}
  \left\langle 
    {\widetilde{\delta}_{lm}}^*(k) \widetilde{\delta}_{l'm'}(k')
  \right\rangle =
  (2\pi)^3 \delta_{ll'} \delta_{mm'} \frac{\delta(k - k')}{k^2} P(k),
\label{eqa13-16}
\end{equation}
where only diagonal elements survive due to the statistical
homogeneity and isotropy of the universe. The density contrast in
redshift space is given by \citep{mat00}
\begin{equation}
  \delta^{\rm (s)}(x(z),\theta,\phi) = 
  D(z)
  \left\{
    b(z) + f(z)
      \left[\frac{\partial}{\partial x} + \alpha(z)\right]
        \frac{\partial}{\partial x} (\triangle + 3K)^{-1}
  \right\}
  \delta(x,\theta,\phi),
\label{eqa3-16-1}
\end{equation}
where $\delta(x,\theta,\phi)$ is the linear density contrast at the
present time, $b(z)$ is the linear bias parameter at redshift $z$,
$f(z)$ is the logarithmic derivative of the growth factor defined by
equation (\ref{eq3-8a}). The linear operator $(\triangle + 3K)^{-1}$
denotes the Green's function of the operator $\triangle + 3K$, and
\begin{equation}
   \alpha(z) = \frac{C_K[x(z)]}{S_K[x(z)]}
   \left\{2 + \frac{d\ln[D(z) f(z) \mPhi(z)]}{d\ln S_K[x(z)]}\right\},
\label{eqa3-16-2}
\end{equation}
where $\mPhi(z)$ is the selection function per comoving volume. The
selection function $\mPhi(z)$ and the redshift distribution $n(z) =
dN(<z)/dz$ of galaxies in the sample with a fixed sky area are related
by $n(z) H(z) \propto \mPhi(z) {S_K}^2(x(z))$. Therefore, the quantity
$\alpha(z)$ is more conveniently expressed in terms of $n(z)$:
\begin{equation}
   \alpha(z) = H(z)
   \frac{d}{dz}\ln[H(z) D(z) f(z) n(z)].
\label{eqa3-16-3}
\end{equation}
In this equation, the term $H d\ln(HDF)/dz$ has the order of the
inverse of the Hubble scale, therefore is negligible, unless the
clustering on Hubble scales is calculated. On the other hand, the term
$H d\ln n/dz$ is the order of inverse of the scale on which the
distribution function $n(z)$ varies. When the correlation function on
such a scale needs to be obtained, the factor $\alpha$ should be
retained. Since the selection function $\mPhi(z)$ usually is a
decreasing function and the volume factor ${S_K}^2(x(z))$ is an
increasing function, the distribution $n(z)=dN/dz$ does not
significantly vary in the useful redshift range of a sample, unless
galaxies near redshift edges are included. Therefore, if the redshift
distribution $n(z)$ is approximately constant in a sample, the factor
$\alpha$ can be neglected. However, when the distribution $dN/dz$
significantly varies in a sample, $\alpha$ must be kept in the
analysis.

It is straightforward to obtain the Green's function since the density
contrast is expanded by the eigenfunction of the Laplacian by equation
(\ref{eqa3-15a}). The correlation function in redshift space is
therefore given by $\xi^{\rm (s)} (z_i, z_j, \theta_{ij}) = \langle
\delta^{\rm (s)}(x(z_i), \theta_i, \phi_i) \delta^{\rm (s)}(x(z_j),
\theta_j, \phi_j) \rangle$, where $\theta_{ij}$ is the angle between
the two directions $(\theta_i,\phi_i)$ and $(\theta_j,\phi_j)$.
Defining the functions,
\begin{equation}
  \xi^{(n)}_l(x) \equiv
  \frac{(-1)^n}{{S_K}^{2n-l}(x)}
  \int\frac{k^2dk}{2\pi^2}
  \frac{\sqrt{N_l(k)}}{k(k^2 - 4K)^n} X_l(k,x) P(k),
\label{eqa3-17}
\end{equation}
we can explicitly represent the linear formula of the two-point
correlation function in redshift space \citep{mat00}:
\begin{equation}
  \xi^{\rm (s)}(z_i,z_j,\theta_{ij}) = 
  b_i b_j D_i D_j
  \sum_{n=0}^2 \sum_{l=0}^{2n}
  c_l^{(n)}(x_i,x_j,\theta_{ij}) \xi_l^{(n)}(x_{ij}),
\label{eqa3-18}
\end{equation}
where we abbreviate $b_i = b(z_i)$, $D_i = D(z_i)$, $x_i = x(z_i)$
etc., $D(z)$ is the linear growth factor normalized by $D(z=0) = 1$,
and $x(z)$ is the comoving distance given by equation (\ref{eq3-4}).
The quantity $x_{ij}$ is the comoving separation of the two-points,
which is obtained from $x_i$, $x_j$, and $\theta_{ij}$ by the
geometric relation,
\begin{eqnarray}
&&
  {S_K}^2(x_{ij}) = {S_K}^2(x_i) + {S_K}^2(x_j)
  - 2 C_K(x_i) C_K(x_j) S_K(x_i) S_K(x_j) \cos\theta_{ij}
\nonumber\\
&&\qquad\qquad\qquad\qquad
  - K {S_K}^2(x_i) {S_K}^2(x_j) \left(1 + \cos^2\theta_{ij}\right).
\label{eqa3-19}
\end{eqnarray}
The coefficients $c_l^{(n)}$ are given by
\begin{eqnarray}
&&
   c^{(0)}_0 = 
   1 + \frac13(\beta_i + \beta_j) +
   \frac{1}{15}\beta_i\beta_j
   \left(1 + 2\cos^2\widetilde{\theta}_{ij}\right),
\label{eqa3-20a}\\
&&
   c^{(1)}_0 = 
   \left[\beta_i + \beta_j +
      \frac{2}{15}\beta_i\beta_j\left(4 + 3\cos\widetilde{\theta}_{ij}\right)
   \right] |K| {S_K}^2(x_{ij}) -
   \frac13 \beta_i \beta_j 
   A_{ij}A_{ji} \cos\widetilde{\theta}_{ij},
\label{eqa3-20b}\\
&&
   c^{(1)}_1 = 
   \beta_i A_{ij}\cos\gamma_{ij} +
   \beta_j A_{ji}\cos\gamma_{ji}
\nonumber\\
&&\qquad\quad + \,
   \frac15 \beta_i \beta_j
   \left[
      A_{ij}
      \left(\cos\gamma_{ij} - 2\cos\gamma_{ji}\cos\widetilde{\theta}_{ij}\right) +
      A_{ji}
      \left(\cos\gamma_{ji} - 2\cos\gamma_{ij}\cos\widetilde{\theta}_{ij}\right)
   \right],
\label{eqa3-20c}\\
&&
   c^{(1)}_2 = 
   \beta_i \left(\cos^2\gamma_{ij} - \frac13 \right) +
   \beta_j \left(\cos^2\gamma_{ji} - \frac13 \right)
\nonumber\\
&&\qquad\quad  - \,
   \frac17 \beta_i \beta_j
   \left[
      \frac23 + \frac43\cos^2\widetilde{\theta}_{ij} -
      \left(\cos^2\gamma_{ij} + \cos^2\gamma_{ji}\right) +
      4\cos\gamma_{ij}\cos\gamma_{ji}\cos\widetilde{\theta}_{ij}
   \right],
\label{eqa3-20d}\\
&&
   c^{(2)}_0 = \beta_i \beta_j
   \left( |K| {S_K}^2(x_{ij}) -
      A_{ij} A_{ji}
   \right) |K| {S_K}^2(x_{ij}),
\label{eqa3-20e}\\
&&
   c^{(2)}_1 = 
   \beta_i \beta_j
   \left(
      A_{ij} \cos\gamma_{ij} +
      A_{ji} \cos\gamma_{ji}
   \right) |K| {S_K}^2(x_{ij}),
\label{eqa3-20f}\\
&&
   c^{(2)}_2 = 
   \frac27 \beta_i\beta_j
   \left[
      \cos^2\widetilde{\theta}_{ij} - \frac23 + 
      \frac92 \left(\cos^2\gamma_{ij} + \cos^2\gamma_{ji}\right) +
      10 \cos\gamma_{ij}\cos\gamma_{ji}\cos\widetilde{\theta}_{ij}
   \right] |K| {S_K}^2(x_{ij})
\nonumber\\
&&\qquad\quad  + \,
   \beta_i\beta_jA_{ij}A_{ji}
   \left(\cos\gamma_{ij}\cos\gamma_{ji} + \frac13 \cos\widetilde{\theta}_{ij}\right),
\label{eqa3-20g}\\
&&
   c^{(2)}_3 = 
   \frac15 \beta_i \beta_j
   \left[
      A_{ij}
      \left(
         5\cos\gamma_{ij}\cos^2\gamma_{ji} - \cos\gamma_{ij} +
         2\cos\gamma_{ji}\cos\widetilde{\theta}_{ij}
      \right)
   \right.
\nonumber\\
&&\qquad\qquad\qquad
   \left. + \,
      A_{ji}
      \left(
         5\cos\gamma_{ji}\cos^2\gamma_{ij} - \cos\gamma_{ji} +
         2\cos\gamma_{ij}\cos\widetilde{\theta}_{ij}
      \right)
   \right],
\label{eqa3-20h}\\
&&
   c^{(2)}_4 = 
   \frac17 \beta_i\beta_j
   \left[
      \frac15 + \frac25\cos^2\widetilde{\theta}_{ij} -
      \left(\cos^2\gamma_{ij} + \cos^2\gamma_{ji}\right) +
      4\cos\gamma_{ij}\cos\gamma_{ji}\cos\widetilde{\theta}_{ij}
   \right.
\nonumber\\
&&\qquad\qquad\qquad  + \,
   \left.
      7\cos^2\gamma_{ij}\cos^2\gamma_{ji}
   \right],
\label{eqa3-20i}
\end{eqnarray}
where the abbreviation $\beta_i =\beta(z_i)$ is employed and $\beta(z)
= f(z)/b(z)$. The quantity $A_{ij}$ is defined by
\begin{equation}
   A_{ij} = S_K(x_{ij}) \alpha(z_i).
\label{eqa3-21}
\end{equation}
The quantity $A_{ji}$ is
similarly defined with the replacement $i\leftrightarrow j$. The
quantity $\gamma_{ij}$ is an angle between the line of sight of $x_i$ and
the direction of the separation $x_{ij}$, which can be obtained by the
equation
\begin{equation}
   \cos\gamma_{ij} =
   \frac{S_K(x_i) C_K(x_j) - C_K(x_i) S_K(x_j)
   \cos\theta_{ij}}{S_K(x_{ij})},
\label{eqa3-22}
\end{equation}
and $\gamma_{ji}$ is similarly defined with the replacement
$i\leftrightarrow j$. Finally, the quantity $\widetilde{\theta}_{ij}$
is defined by
\begin{equation}
  \cos\widetilde{\theta}_{ij} = 
  \frac{C_K(x_i) C_K(x_j) \cos\theta_{ij} + K S_K(x_i) S_K(x_j)}
    {C_K(x_i) C_K(x_j) + K S_K(x_i) S_K(x_j) \cos\theta_{ij}} =
  \frac{\sin\gamma_{ij} \sin\gamma_{ji}}{C_K(x_{ij})} -
  \cos\gamma_{ij} \cos\gamma_{ji}.
\label{eqa3-23}
\end{equation}
The matrix $\widetilde{\theta}_{ij}$ is symmetric, while $A_{ij}$ and
$\gamma_{ij}$ are not symmetric.

Although the expression of the two-point correlation function
(\ref{eqa3-18}) is somewhat tedius, the numerical calculation is
straightforward. Especially, once the single-variable functions
$f(z)$, $D(z)$, $x(z)$, $b(z)$, $\mPhi(z)$, $\xi_l^{(n)}(x)$ are
calculated and tabulated beforehand, then the evaluation of the
correlation function does not require any further numerical
integrations. This property is the essential part for the fast
computation of the correlation matrix we have developed in this paper.


\section{Window Functions and the Epanechnikov Kernels
\label{app3}}

In this appendix, window functions for a series of kernel functions,
i.e., Epanechnikov kernels are explicitly given.

\subsection{The Three-dimensional Case}

First we consider the three-dimensional space. The window function
is the 3-dimensional Fourier transform of the kernel function:
\begin{equation}
  W(kR) =
  \int d^3x W_R(x) e^{-i\sbfm{k}\cdot\sbfm{x}} =
  4\pi \int_0^\infty x^2 dx W_R(x) j_0(kx)
\label{eqa4-1}
\end{equation}
There are two popular spherical kernels, i.e., the top-hat kernel,
\begin{equation}
  W^{\rm (T)}_R(x) = \frac{3}{4\pi R^3} \mTheta(R-x), \qquad
  W^{\rm (T)}(kR) = \frac{3 j_1(kR)}{kR},
\label{eqa4-2}
\end{equation}
where
\begin{equation}
  \mTheta(x) = 
  \left\{
  \begin{array}{ll}
    1, & (x \geq 0), \\
    0, & (x < 0),
  \end{array}
  \right.
\label{eqa4-3}
\end{equation}
is the Heaviside step function, and the Gaussian kernel,
\begin{equation}
  W^{\rm (G)}_R(x) = \frac{1}{(2\pi)^{3/2} R^3} e^{-x^2/(2R^2)}, \qquad
  W^{\rm (G)}(kR) = e^{-(kR)^2/2}.
\label{eqa4-4}
\end{equation}
One of the advantages of the top-hat kernel is that it has finite
support. It is a disadvantage that the Gaussian kernel extends to
infinite volume, which should be truncated in actual analyses.
However, the Gaussian kernel has an advantage that the window function
$W^{\rm (G)}(kR)$ drops off exponentially for large wavenumbers $k$,
and thus the numerical integration with this factor is very stable.
The function $W^{\rm (T)}$ is an oscillating function and the envelope
drops off as $(kR)^{-2}$, which converges much slower than the
Gaussian window.

The slow drop-off of the Fourier transform of the top-hat kernel comes
from the sharp discontinuity of the kernel at the edge, $x=R$.
Accordingly, it is sometimes advantageous to use a kernel that is
continuous at the edge, and at the same time, has finite support. One
of such kernels is the Epanechnikov kernel,
\begin{equation}
  W^{\rm (E)}_R(x) =
  \frac{15}{8\pi R^3}
  \left(1 - \frac{x^2}{R^2}\right) \mTheta(R-x), \qquad
  W^{\rm (E)}(kR) = \frac{15 j_2(kR)}{(kR)^2},
\label{eqa4-5}
\end{equation}
which has a parabolic profile and the envelope of the Fourier
transform drops off as $(kR)^{-3}$. While the Epanechnikov kernel is
contiuous on the edge, the radial derivative is not. The Epanechnikov
kernel can be generalized such that the radial derivatives on the edge
are also continous. The generalized kernel is called $m$-weight
Epanechnikov kernel, which is given by
\begin{equation}
  W^{({\rm E}m)}_R(x) =
  \frac{(2m+3)!!}{2^{m+2} m! \pi R^3}
  \left(1 - \frac{x^2}{R^2}\right)^m \mTheta(R-x), \quad
  W^{({\rm E}m)}(kR) = \frac{(2m+3)!! j_{m+1}(kR)}{(kR)^{m+1}}.
\label{eqa4-6}
\end{equation}
The original Epanechnikov kernel is the $1$-weight kernel, $W^{({\rm
E}1)}_R$. Top-hat kernel is also from this sequence with $m=0$. The
$m$-weight Epanechnikov kernel is countinous up to $(m-1)$ derivatives
on the edge. The envelope of the Fourier transform drops off as
$(kR)^{-m-2}$, so that the convergence is faster for higher orders.
The effective width of the kernel is smaller than $R$, especially for
higher weights $m$. In this respect, the size of the kernels is more
conveniently represented by the ``variance'' of the kernels,
\begin{equation}
  {\cal R} \equiv
  \left[
    \int d^3x W_R^{({\rm E}m)}(x) x^2
  \right]^{1/2} = 
  \left(\frac{3}{2m+5}\right)^{1/2} R.
\label{eqa4-7}
\end{equation}
For the top-hat kernel, the smoothing radius $R_{\rm T}$ corresponds
to $R_{\rm T} = (5/3)^{1/2}{\cal R}$. In the limit $m \rightarrow
\infty$, with ${\cal R}$ fixed, the $m$-weight Epanechnikov kernel
reduces to the Gaussian kernel. In fact,
\begin{equation}
  W_R^{({\rm E}m)}(x)
  \ \mathop{\longrightarrow}_{\stackrel{m\rightarrow\infty}{
  {\cal R}\ \mbox{\scriptsize fixed}}}
  \ W^{\rm (G)}_{R_{\rm G}}(x),
\label{eqa4-8}
\end{equation}
where the Gaussian smoothing length $R_{\rm G}$ corresponds to
$R_{\rm G} = {\cal R}/\sqrt{3}$. Therefore, the generalized
Epanechnikov kernels contain both the top-hat kernel and the Gaussian
kernels at $m=0$ and $m=\infty$, respectively.

\subsection{The Two-dimensional Case}

Next we summarize the case for the two-dimensional space. The window
function is given by
\begin{equation}
  W(l \theta_{\rm s}) =
  \int d^2\theta W_{\theta_{\rm s}}(\theta)
     e^{-i\sbfm{l}\cdot\sbfm{\theta}} =
  2\pi \int_0^\infty  \theta d\theta W_{\theta_{\rm s}}(\theta) J_0(l\theta).
\label{eqa4-11}
\end{equation}
The top-hat kernel and its window function are 
\begin{equation}
  W^{\rm (T)}_{\theta_{\rm s}}(\theta)
  = \frac{1}{\pi {\theta_{\rm s}}^2} \mTheta(\theta_{\rm s}-\theta), \qquad
  W^{\rm (T)}(l\theta_{\rm s}) = 
  \frac{2 J_1(l\theta_{\rm s})}{l\theta_{\rm s}},
\label{eqa4-12}
\end{equation}
and Gaussian counterparts are
\begin{equation}
  W^{\rm (G)}_{\theta_{\rm s}}(\theta) =
  \frac{1}{2\pi {\theta_{\rm s}}^2} e^{-\theta^2/(2{\theta_{\rm s}}^2)},
  \qquad
  W^{\rm (G)}(l \theta_{\rm s}) = e^{-l^2 {\theta_{\rm s}}^2/2}.
\label{eqa4-13}
\end{equation}

The Epanechnikov kernel in 2-dimension is given by
\begin{equation}
  W^{\rm (E)}_{\theta_{\rm s}}(\theta) =
  \frac{2}{\pi {\theta_{\rm s}}^2}
  \left(1 - \frac{\theta^2}{{\theta_{\rm s}}^2}\right)
    \mTheta(\theta_{\rm s}-\theta), \qquad
  W^{\rm (E)}(l\theta_{\rm s}) =
  \frac{8 J_2(l\theta_{\rm s})}{(l\theta_{\rm s})^2}.
\label{eqa4-14}
\end{equation}
which has a parabolic profile and the envelope of the Fourier
transform drops off as $(l\theta_{\rm s})^{-5/2}$. The $m$-weight
Epanechnikov kernel is derived as
\begin{equation}
  W^{({\rm E}m)}_{\theta_{\rm s}}(\theta) =
  \frac{m+1}{\pi {\theta_{\rm s}}^2}
  \left(1 - \frac{\theta^2}{{\theta_{\rm s}}^2}\right)^m
  \mTheta(\theta_{\rm s}-\theta), \quad
  W^{({\rm E}m)}(l\theta_{\rm s}) = 
  \frac{2^{m+1} (m+1)! J_{m+1}(l\theta_{\rm s})}{(l\theta_{\rm s})^{m+1}}.
\label{eqa4-15}
\end{equation}
The envelope of the Fourier transform drops off as $(kR)^{-m-3/2}$, so
that the convergence is faster for higher weights. The effective width
of the kernel is smaller than $R$, especially for higher weights $m$.
The size of the kernels is alternatively represented by the
``variance'' of the kernels,
\begin{equation}
  \vartheta_{\rm s} \equiv
  \left[
    \int d^2\theta W_{\theta_{\rm s}}^{({\rm E}m)}(\theta) \theta^2
  \right]^{1/2} = 
  \frac{\theta_{\rm s}}{\sqrt{m+2}}
\label{eqa4-16}
\end{equation}
For the top-hat kernel, the smoothing radius $\theta_{\rm T}$
corresponds to $\theta_{\rm T} = \sqrt{2}\,\vartheta_{\rm s}$. In the
limit $m \rightarrow \infty$, with $\vartheta_{\rm s}$ fixed, the
$m$-weight Epanechnikov kernel reduces to the Gaussian kernel. In
fact,
\begin{equation}
  W_{\theta_{\rm s}}^{({\rm E}m)}(\theta)
  \ \mathop{\longrightarrow}_{\stackrel{m\rightarrow\infty}{
  \vartheta_{\rm s}\ \mbox{\scriptsize fixed}}}
  \ W^{\rm (G)}_{\theta_{\rm G}}(x),
\label{eqa4-17}
\end{equation}
where the Gaussian smoothing length $\theta_{\rm G}$ corresponds to
$\theta_{\rm G} = \vartheta_{\rm s}/\sqrt{2}$.



\newpage

\end{document}